\documentclass{article}
\usepackage[a4paper, portrait, margin=1.1811in]{geometry}
\usepackage[english]{babel}
\usepackage[utf8]{inputenc}
\usepackage[T1]{fontenc}
\usepackage{helvet}
\usepackage{etoolbox}
\usepackage{graphicx}
\usepackage{titlesec}
\usepackage{multicol}
\usepackage{caption}
\usepackage{booktabs}
\usepackage{xcolor} 
\usepackage[colorlinks, citecolor=cyan]{hyperref}
\usepackage{caption}
\graphicspath{ {./images/} }
\usepackage{scrextend}
\usepackage{fancyhdr}
\usepackage{caption}
\usepackage{subcaption}
\usepackage{graphicx}
\newcounter{lemma}

\newcounter{theorem}

\usepackage{amsfonts}
\usepackage{amsmath}

\fancypagestyle{plain}{
	\fancyhf{}

}

\makeatletter
\patchcmd{\@maketitle}{\LARGE \@title}{\fontsize{16}{19.2}\selectfont\@title}{}{}
\makeatother

\newcommand{\argmin}{\mathop{\mathrm{arg\,min}}}  

\usepackage{authblk}

\newsavebox\affbox

\author[1*,3]{\textbf{Gerhard Hellstern}}
\author[2]{\textbf{Vanessa Dehn}}
\author[3]{\textbf{Martin Zaefferer}}

\affil[1] {Center of Finance, DHBW Stuttgart, Herdweg  29, D-70174 Stuttgart, Germany}

\affil[2]{ Fraunhofer-Institut f\"ur Angewandte Festk\"orperphysik IAF, Tullastr. 72, D-79108 Freiburg, Germany}

\affil[3] {DHBW Ravensburg, Marienplatz 2, D-88212 Ravensburg, Germany}

\titlespacing\section{0pt}{12pt plus 4pt minus 2pt}{0pt plus 2pt minus 2pt}
\titlespacing\subsection{12pt}{12pt plus 4pt minus 2pt}{0pt plus 2pt minus 2pt}
\titlespacing\subsubsection{12pt}{12pt plus 4pt minus 2pt}{0pt plus 2pt minus 2pt}

\titleformat{\section}{\normalfont\fontsize{10}{15}\bfseries}{\thesection.}{1em}{}
\titleformat{\subsection}{\normalfont\fontsize{10}{15}\bfseries}{\thesubsection.}{1em}{}
\titleformat{\subsubsection}{\normalfont\fontsize{10}{15}\bfseries}{\thesubsubsection.}{1em}{}

\titleformat{\author}{\normalfont\fontsize{10}{15}\bfseries}{\thesection}{1em}{}

\title{\textbf{\huge Quantum computer based Feature Selection in Machine Learning}}
\date{}    

\begin{document}

\pagestyle{headings}	
\newpage
\setcounter{page}{1}
\renewcommand{\thepage}{\arabic{page}}

\captionsetup[figure]{labelfont={bf},labelformat={default},labelsep=period,name={Figure }}	\captionsetup[table]{labelfont={bf},labelformat={default},labelsep=period,name={Table }}
\setlength{\parskip}{0.5em}
	
\maketitle
	
\noindent\rule{15cm}{0.5pt}
	\abstract
        The problem of selecting an appropriate number of features in supervised learning problems is investigated in this paper. Starting with common  methods in machine learning, we treat the feature selection task as a quadratic unconstrained optimization problem (QUBO), which can be tackled with classical numerical methods as well as within a quantum computing framework. We compare the different results in small-sized problem setups. According to the results of our study, whether the QUBO method outperforms other feature selection methods depends on the data set. In an extension to a larger data set with 27 features, we compare the convergence behavior of the QUBO methods via quantum computing with classical stochastic optimization methods. Due to persisting error rates, the classical stochastic optimization methods are still superior.

\section{Introduction and literature review}
When using supervised machine learning to tackle real-world problems one is faced with the task of selecting the "right" features for the learning procedure. When too few features are chosen, this may lead to poor results for the algorithms. When too many features are selected, the training time is negatively affected, which may lead to unstable results. Moreover, the more features are used, the more effort is required to maintain and control these features for a deployed  machine learning model.
Just testing all possible combinations of features with a brute-force approach is usually not feasible, since the number of combinations grows exponentially with the number of features.

In the past, different methods have been proposed to tackle this problem. Brown et al. provide a compilation of several approaches where {\it mutual information} is used as a dependency measure, introduced within an information-theoretic framework \cite{Brown.2012}. An extensive survey of feature selection methods that are applied to different data sets to demonstrate their behavior is given in  \cite{CHANDRASHEKAR201416} and \cite{Venkateswara2015}. 

The idea of using quantum computers for solving this problem is introduced in \cite{Milne2017}: There, the feature selection problem has been reformulated as a quadratic unconstrained optimization problem (QUBO). In addition, this problem was solved via classical optimization methods as well as with a quantum annealing approach. The different methods were applied to a  well-known data set, the "German Credit Data" of UC Irvine \cite{Dua:2019}.

The annealing approach of solving the QUBO problem for feature selection was also employed in \cite{Nembrini.2021}, where the authors investigate feature selection for recommender systems. With the same approach, other data sets were explored in \cite{Dacrema.22}. Results that combine results obtained with annealing methods as well as with a gate-based approach to quantum computing are reported in \cite{Muecke2023}. An approach that examines an unconstrained black box binary optimization and applies it to feature selection is presented in \cite{Zoufal2023}. Finally, \cite{Turati_2022} compares different algorithms (annealing, gate-based and classical) to solve the QUBO problem of feature selection in a small-size setup.

In this paper, we report on the following approach to investigate the feature selection problem: Using small-sized data sets we first compare well-known machine-learning methods for feature selection with the brute-force method. 

After reformulating the feature selection problem as a QUBO, hereby allowing different dependency measures between the features and also between the features and the target, we investigate  the stability of the solutions when using different values of the undetermined weighting parameter with classical optimization methods. Since for this problem size the exact solution for the features selection problem can be determined, we investigate to what extent this solution can be achieved on the one hand with the commonly used feature selection methods and on the other hand with the way via QUBO optimization. 

In order to solve the QUBO with the gate-based approach of quantum computing, we use the QAOA- and the VQE-ansatz. 
In addition to using the out-of-the-box QAOA algorithm of IBM, we compare the results with a customized solution scheme for QAOA. 
Hereby it can be shown whether and to what extent the QAOA algorithm is a feasible way to solve the QUBO problem and how strong the goodness of the solutions depends on the QAOA depth as well as on the dependence measures used. 

After discussing our results on small-sized data sets, we scale our methods up to 27 features (i.e., 27 qubits) and report optimization results on different physical quantum computers of IBM. For this setup, we also compare the results to several classical stochastic optimization methods. At this point, we investigate, on the one hand, whether 27 qubits on a real quantum computer can yield any meaningful results beyond random noise, given its error rates. On the other hand, we focus on comparing the convergence of the QAOA algorithm with a selection of classical stochastic optimization methods.

\section{Feature Selection in Supervised Learning}
\subsection{Machine Learning approaches}

In the following, we restrict ourselves to binary classification as one of the most important topics in supervised learning: 

Given are a real  $m \times n$  matrix $X$ consisting of $n$ features and $m$ training examples and a target vector $y$; $y$ is of dimension $m$ and each element $y_i \in \left\lbrace 0,1 \right\rbrace $. A classification model is a mapping $f: X \in A \rightarrow B$, i.e. from an input space $A$ to an output space $B$. In order to find the best model $f$, we use a loss function $g: A \times B \rightarrow \mathbb{R}$ that has to be minimized. In the case of a binary classification problem, the binary cross-entropy is mostly used as a loss function.
For a thorough discussion of machine learning, see e.g. \cite{Bishop2006}.

Given a training set $(X,y)$, the following problem arises in practice: How many and which features should be included in the training of a particular model in order to do the best job? Naively, one would expect that considering all features is naturally the best choice. However, there are several reasons why this is in general not the case:

\begin{itemize}
	\item Features with a high dependence on each other can lead to an unstable training process.
	
	\item Each feature used by an algorithm must be prepared and pre-processed and controlled in a productive environment. So, from an economical point of view, one should not consider more  features than necessary.

    \item Models for data with more features require more computing time.
    
	\item It may happen, that some of the features have no predictive power at all. Identifying and removing them can save costs.
\end{itemize}	

Selecting the optimal number of features by brute force, i.e. testing all possible combinations, is in general not feasible since the combinatorial possibilities grow exponentially with the number of features. Therefore, several methods for feature selection have been proposed and applied in the past:

\begin{itemize}
	\item Using a penalty term $\alpha \sum_{i=1}^{n} \mid w_i \mid $ with an adjustable hyper-parameter $\alpha$ in the cost function of the training algorithm effectively reduces the number of features. This is called LASSO \cite{LASSO} in the machine learning context. 
	
	\item The recursive feature elimination procedure (RFE) starts with all possible features and then reduces the number of features according to their importance one by one in a greedy manner. Instead of comparing all combinations of features, the number of models trained and compared corresponds to the number of features.
	
	\item One may also use unsupervised methods, e.g., principal component analysis (PCA), to reduce the number of features in the model. However, this is done independently by the target and can therefore not take into account which features may have high predictive power. Also, PCA and similar methods generate a (lesser number of) new features from the original features, rather than merely selecting. Hence, the original features would still have to be available when deploying this approach in practice.
\end{itemize}	

\subsection{QUBO-formulation of the problem}
Here, following the suggestion of \cite{Milne2017}, we propose and explore a different approach for feature selection: Let $\rho_{i,j}$ be a dependency measure between column $i$ and column $j$ of the matrix $X$ and let $\rho_{i,Y}$ be the dependence between column $i$ and the target vector. Later, we will specify which dependency measures are used. 

In addition, let $z_i \in \{0,1 \}$ be a binary variable, indicating if column $i$ should be selected ($z_i=1$) in the model or not ($z_i=0$).
Intuitively, features should be selected so that the dependence 
\begin{equation}
	\sum_{i,j=1, i\neq j}^{n} z_i z_j \mid \rho_{i,j}\mid, 
\end{equation}
between them is minimized,
whereas the dependence between the features and the target vector 
\begin{equation}
	\sum_{i=1}^{n} z_i \mid \rho_{i,Y}\mid.
\end{equation}
is maximized. 
Note, that we take the absolute values of the dependency measures since for the dependence measures considered in the following (e.g. correlation) a high negative dependence is as relevant as a high positive dependence.  

Putting both criteria together, we get the following optimization (here: minimization) problem:
\begin{equation}
	h(\boldmath z)=- \left[ \phi 	\sum_{i=1}^{n} z_i \mid \rho_{i,Y}\mid -(1-\phi) \sum_{i,j=1, i\neq j}^{n} z_i z_j \mid \rho_{i,j}\mid \right] 
\label{QUBO_0}	
\end{equation}
The parameter $\phi \in [0,1] $ has to be considered as a hyper-parameter and governs the weighting between the two conditions. By tuning $\phi$ one can decide which of the two conditions is more important. Using the fact that $z_i$ is a binary variable, i.e. $z_i z_i=z_i$, the above equation can be transformed to
\begin{equation}	
	h(\boldmath z)=- \left[ {\boldmath z^T} Q {\boldmath z}   \right]
\end{equation}
and the solution to our problem is given by 
\begin{equation}
	{\boldmath z}^*= 
 \argmin_{z}
 \left[ {h(\boldmath z)}   \right]
 \label{QUBO}
\end{equation}

Such a problem, known as quadratic unconstrained binary optimization (QUBO) problem can be treated with classical numerical methods but is also suited for a solution with a quantum computer.

\begin{figure}
	\centering	
	\includegraphics[width=0.7\linewidth]{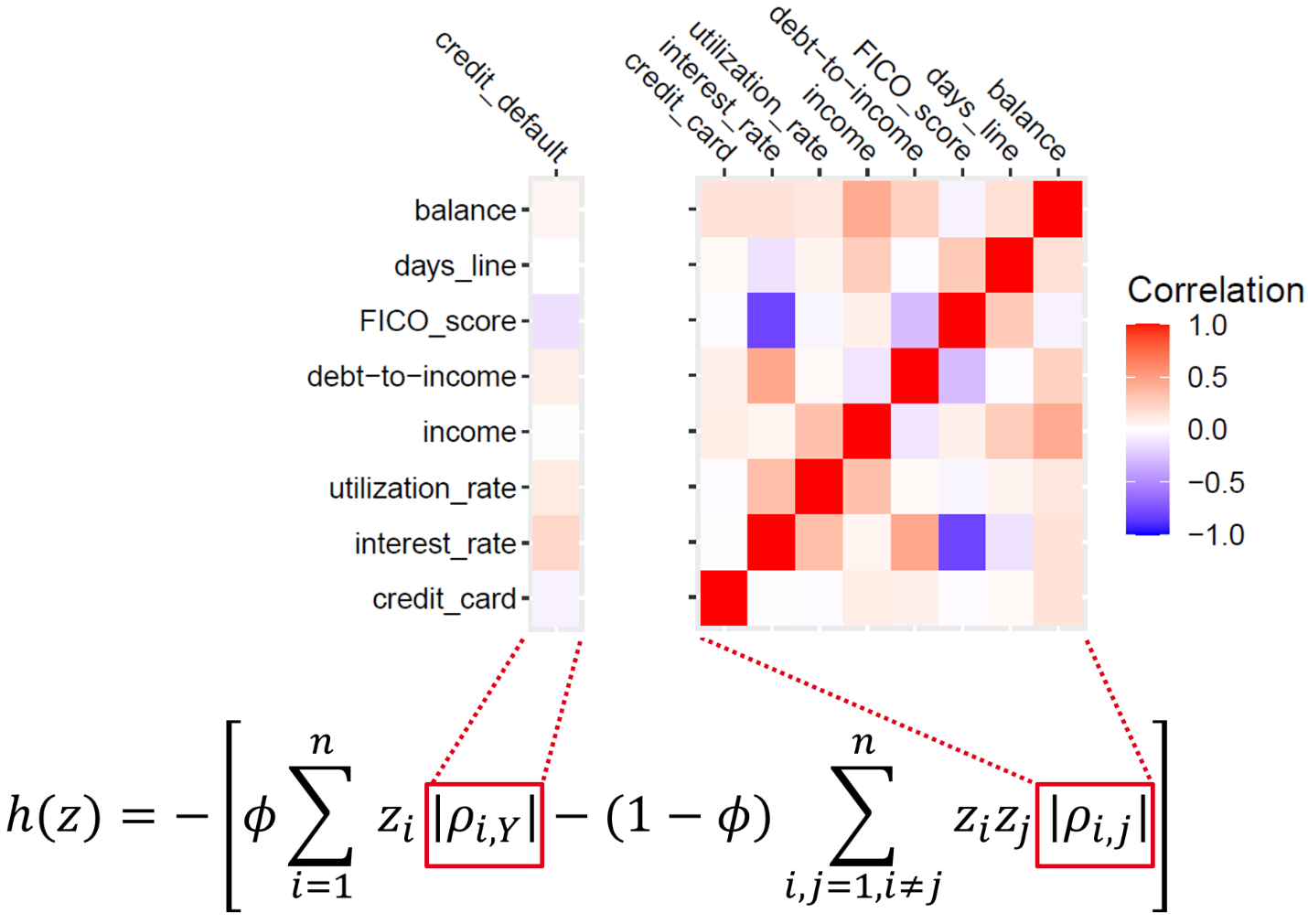}
	\caption[]{Illustration of the objective function for lending club data.}
\label{fig:objective}
\end{figure}

Up to now the dependency measure is not yet determined. The most natural choice would be to use correlation as a linear measure, ${\rho }^{Correl}$, as shown in Fig.~\ref{fig:objective}. Alternatively, one could choose the rank correlation. 

Other possible dependency measures are the mutual information $\rho^{MI}$ which has been discussed in \cite{Brown.2012}, the univariate ROC-value $\rho^{ROC}$ and the Anova F-Statistic $\rho^{Anova}$. All these dependency measures could be used either between the features or between the features and the target variable. 

In the following, for the inter-feature dependence we restrict ourselves to correlation and the mutual information measure. For the dependence on the target, we tested all measures. In the following, we will denote the selected dependence by the tuple $(\rho_{Feature}^X, \rho_{Target}^Y)$ where $X \in \{Correl, MI \}$ and 
$Y \in \{Correl, MI,ROC,Anova\}$.

\section{Classical feature selection for a small-sized data set}
\subsection{Machine Learning method}

In this chapter, a comparison of the described feature selection methods is performed. To that end, we consider two data sets: a) a subset of Kaggle's lending club data \cite{Kaggle2022} and b) the Wisconsin breast cancer data \cite{Dua:2019}. In the case of a) we use 8 features and 1000 observations; in the case of b) we have 10 features and 569 observations in the data set. For both data sets we seek a binary classification model with the "optimal" number of features. To keep the analysis transparent we use logistic regression as the machine learning model and use the whole data set for training. We are aware, that this may lead to overfitting, however, logistic regression, in comparison to more sophisticated models, proves often to be robust against overfitting (as long as the model term is restricted to low-order polynomials).

For comparing models with different input features, we use the overall accuracy (ACC) of the model as well as the area under curve (AUROC) as performance measures.
Accuracy is the proportion of correctly classified cases by the total number of all data points. AUROC is the area under the ROC curve which is constructed by plotting the true positive rate (TPR) against 1-FPR, i.e. false positive rate (FPR). Intuitively AUROC measures the ability of the model to separate the two classes which is desirable in many use cases.

To find the optimal number of features, we use these two measures as relevant criteria. Since the accuracy of a model may provide misleading results in case of imbalanced data sets, we chose the AUROC as the measure to optimize.

For a small number of features $n$, it is possible to determine the optimal feature selection by brute force, i.e. simply trying all possible combinations of features and then training the corresponding model.
The determined optimal choice is  compared to the results obtained with the RFE- and the LASSO-method. For LASSO we used cross-validation to find the optimal value of $\alpha$. 

In the case of the lending club data, we observed the results shown in Table~\ref{tab:resultsLending}.
\begin{table}
    \centering
    \caption{Lending club data: performance of the logistic regression model for different feature selection approaches.}
    \label{tab:resultsLending}
\begin{tabular}{|c|c|c|c|}
	\hline
	& AUROC & Accuracy & Number of features \\
	\hline
	All features & 0.6651 & 0.83 & 8 \\
	\hline
	Best choice of features & {\bf 0.6682} & 0.83 &  6 \\
	\hline
	Selection with RFE & 0.6593 & 0.83 & 1 \\
	\hline
	Selection with LASSO & 0.6681 & 0.83 &  3\\
	\hline
\end{tabular}
\end{table}
While the accuracy is robust against different choices of features, this is not the case for the AUROC. The best choice of features (obtained with brute force) can not be obtained via RFE or LASSO: 
The optimal choice is six out of eight features; RFE recommends keeping only one feature while LASSO recommends keeping three features out of eight.

In the case of the breast cancer data, the results are shown in Table~\ref{tab:resultsBreast}.
\begin{table}
    \centering
    \caption{Breast cancer data: performance of the logistic regression model for different feature selection approaches.}
    \label{tab:resultsBreast}
\begin{tabular}{|c|c|c|c|}
	\hline
	& AUROC & Accuracy & Number of features \\
	\hline
	All features & 0.9844 & 0.9297 & 10\\
	\hline
	Best choice of features & {\bf 0.9863} & 0.9297 & 5 \\
	\hline
	Selection with RFE & 0.9839 & 0.9297 & 6 \\
	\hline
	Selection with LASSO & 0.9845 & 0.9262 &  3\\
	\hline
\end{tabular}
\end{table}
The optimal choice is five out of ten features; RFE recommends to keep six features and LASSO recommends to keep three features out of eight. Again, the best choice of features is not obtained with RFE or LASSO.

The results obtained so far should be considered as a benchmark and it has to be shown, whether feature selection via optimization as described above, is able to compete with these benchmarks.

\subsection{Optimization methods}

To answer the previous question, we treat the feature selection task as an optimization problem. Again, due to the low dimensionality of the two problems, the exact solution of the optimization, c.f. equation (\ref{QUBO}), can be obtained by brute force. 

However, it is not obvious, which value of $\phi$ should be chosen to obtain the best solution. To tackle this problem and to explore the dependence of the results on $\phi$, we solve the optimization problem for different values of $\phi$ and compare the results to each other. Here, we also consider different combinations of dependency measures - between the features, but also between each feature and the target vector. 

\begin{figure}
  \centering
  \begin{minipage}{.40\linewidth}
    \centering
    \subcaptionbox{Lending club: AUROC as a function of $\phi$ for different dependency measures (1)}
      {\includegraphics[width=\linewidth]{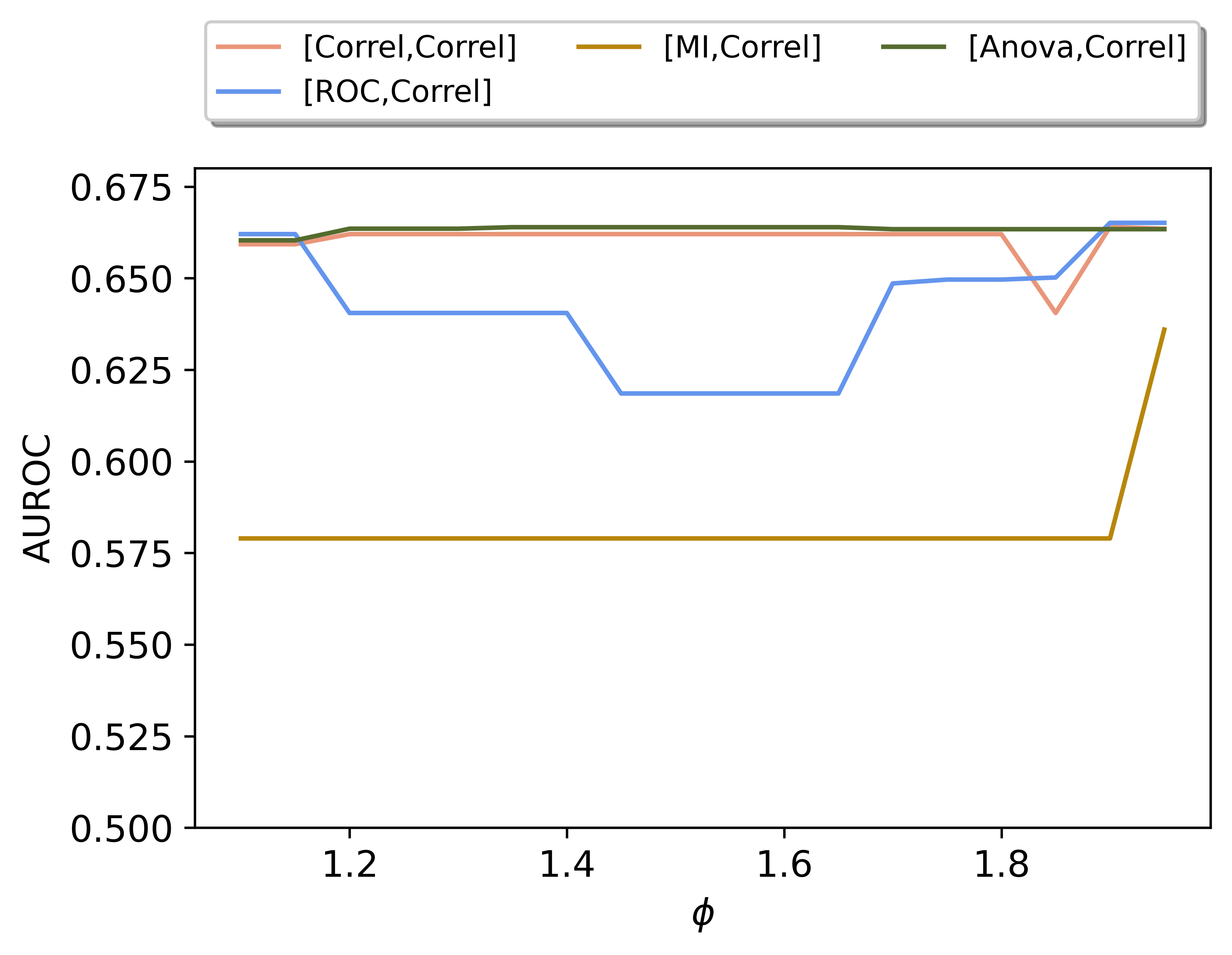}}

    \subcaptionbox{Lending club: AUROC as a function of $\phi$ for different dependency measures (2)}
      {\includegraphics[width=\linewidth]
      {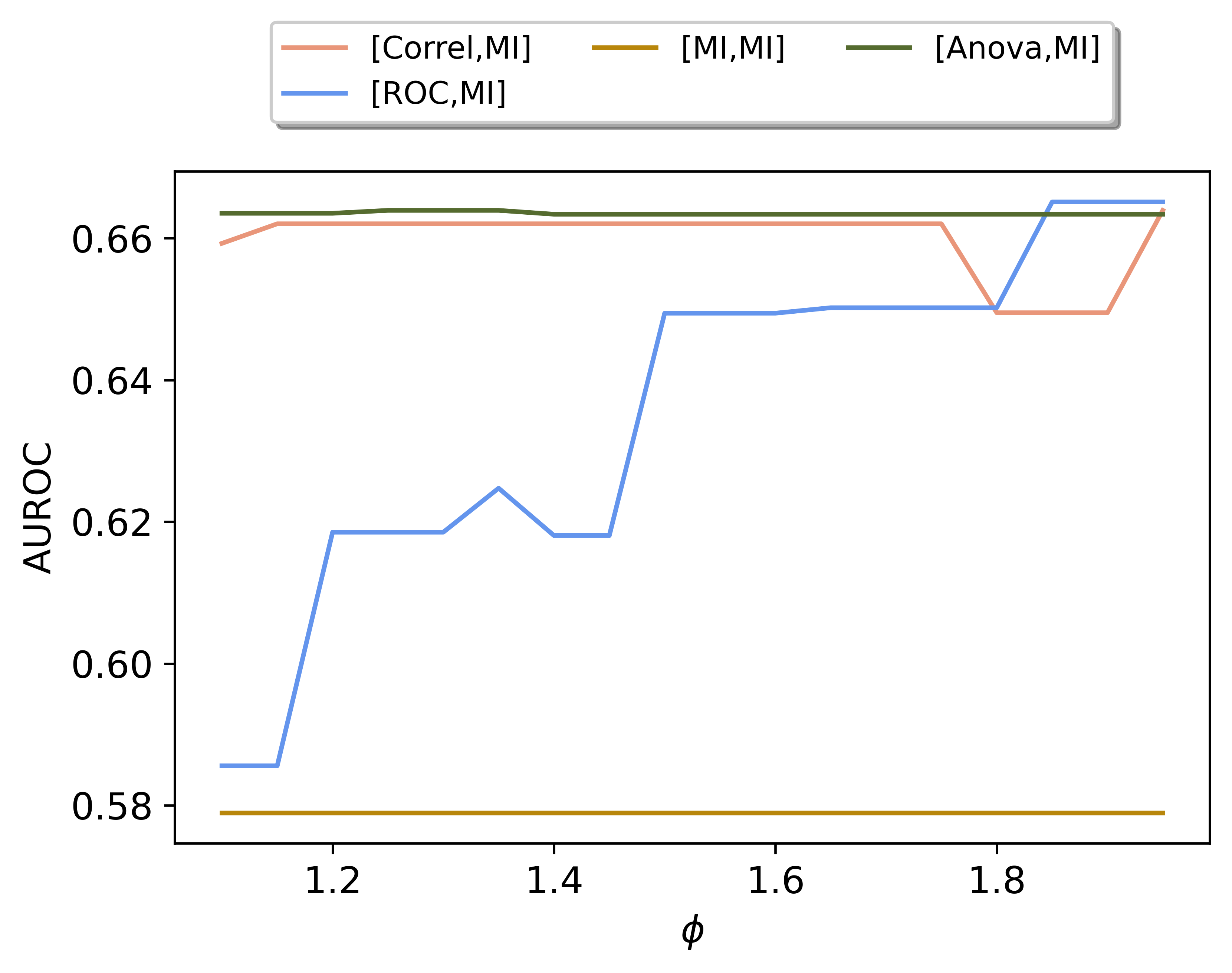}}

    \subcaptionbox{Lending club: AUROC as a function of $\phi$; best obtainable result}
      {\includegraphics[width=\linewidth]
      {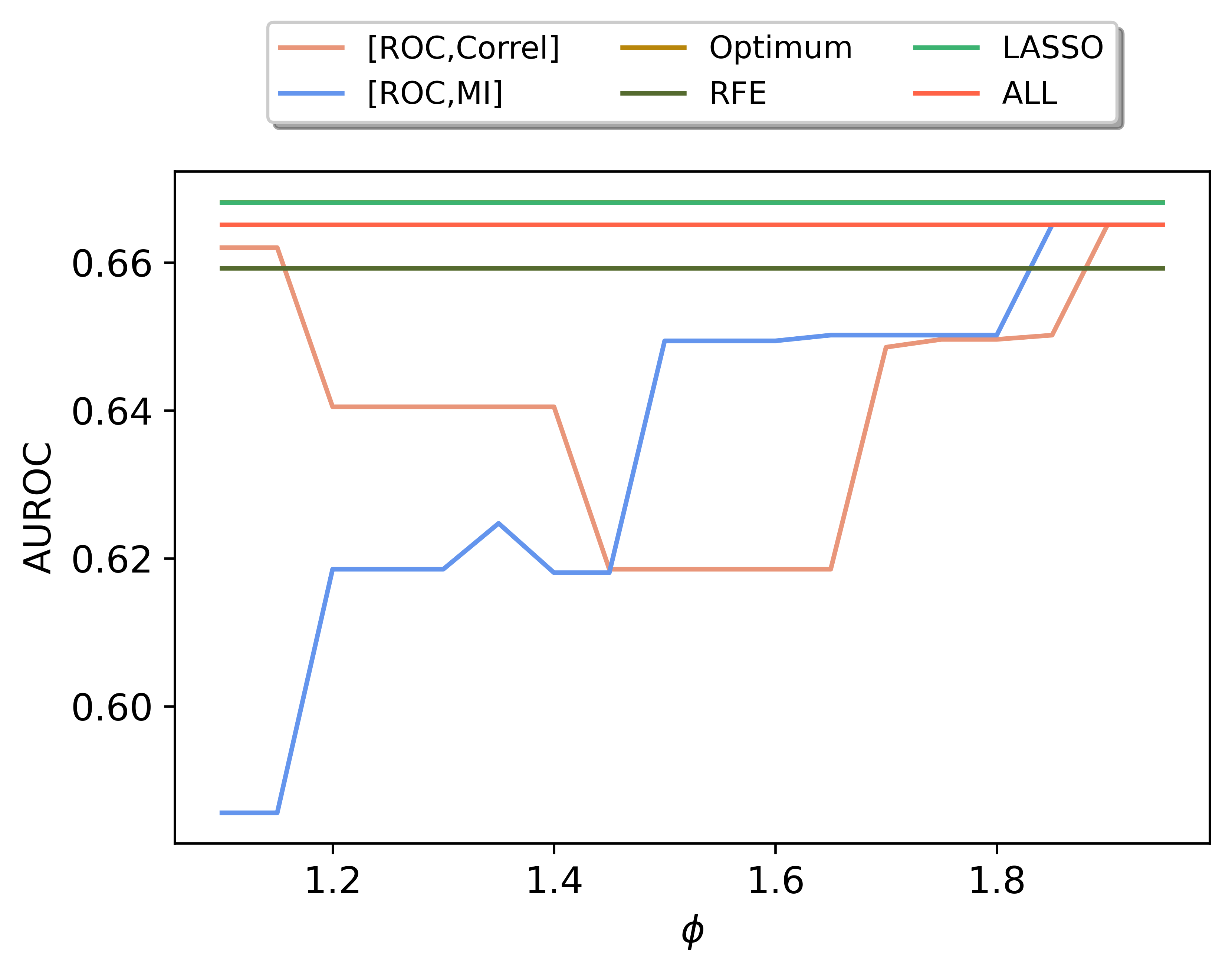}}

\caption{Lending club data \label{fig:bestphi}}
  \end{minipage}\quad
  \begin{minipage}{.40\linewidth}
    \centering
    \subcaptionbox{Breast cancer: AUROC as a function of $\phi$ for different dependency measures (1)}
      {\includegraphics[width=\linewidth]
      {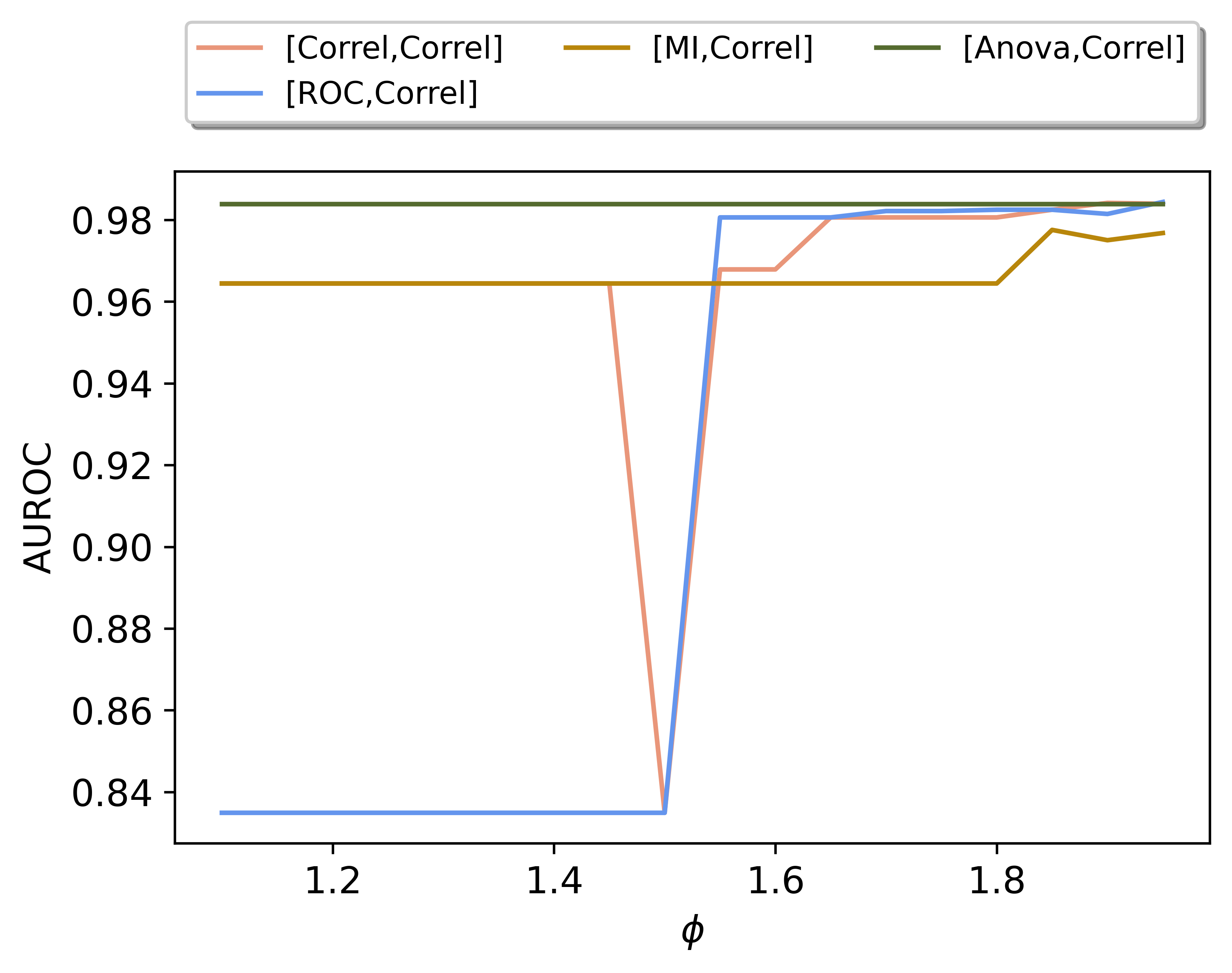}}

    \subcaptionbox{Breast cancer: AUROC as a function of $\phi$ for different dependency measures (2)}
    {\includegraphics[width=\linewidth]{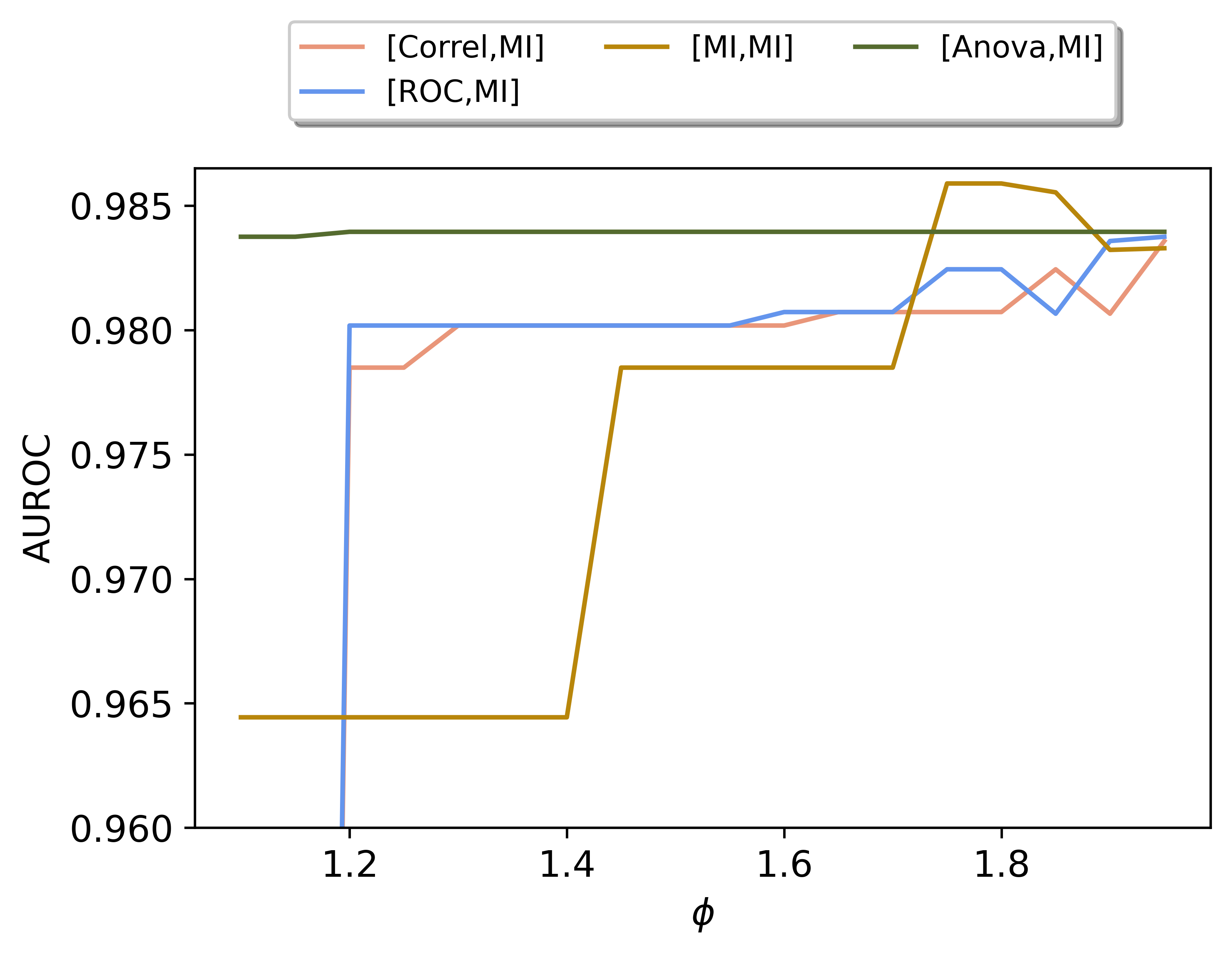}}

    \subcaptionbox{Breast cancer: AUROC as a function of $\phi$; best obtainable result}
      {\includegraphics[width=\linewidth]{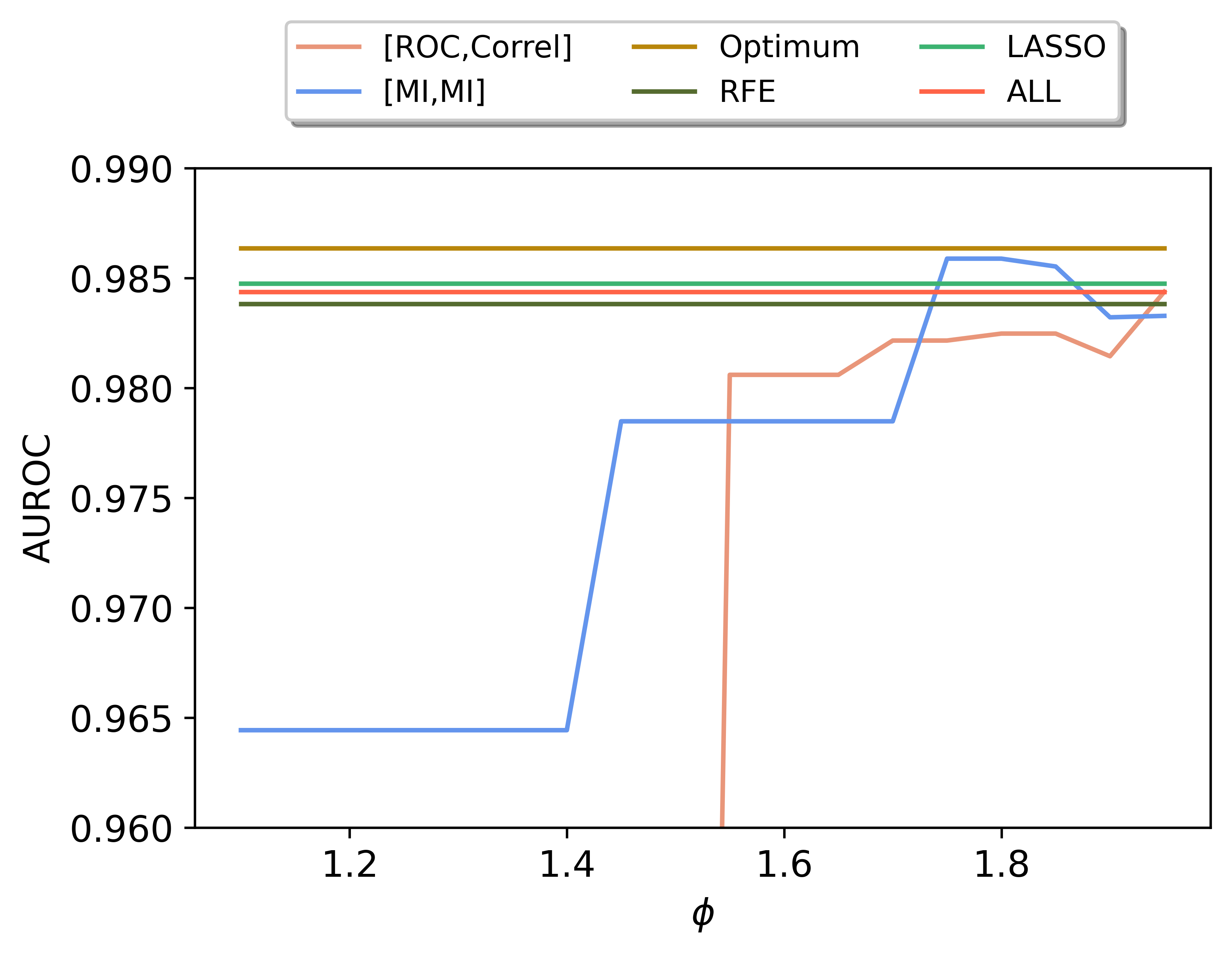}}

\caption{Breast cancer data \label{fig:bcbestphi}}
  \end{minipage}

  \bigskip
\end{figure}

In Fig.~\ref{fig:bestphi} and Fig.~\ref{fig:bcbestphi} the results for the lending club data are shown. Several findings can be observed in these figures:

\begin{itemize}

\item Using mutual information as the dependence measure between features, irrespective of the measure to the target, leads to the worst AUROC values.

\item When using ANOVA as the dependency measure, we observe that the AUROC is largely independent of $\phi$. 

\item In most cases, we see that the AUROC depends on $\phi$; the AUROC tends to increase with increasing values of $\phi$.

\end{itemize}

In all of our experiments, we find the highest value of the AUROC when using the ROC as the dependency measure between the features and the correlation or mutual information between each feature and the target. In addition, the $\ phi$ value should be chosen larger than 0.85. Note, $\phi=1$ is the case where only the  dependence between features and target is considered. The $\phi-$dependence is displayed in Fig. \ref{fig:bestphi}. Here, we show for comparison - as a straight line - the best AUROC of the best model (obtained by brute force), the AUROC of the RFE- and LASSO models. It can be seen, that RFE is outperformed by our proposed optimization approach with a suitable choice of dependency measure. The LASSO result is only slightly worse than the best possible. The best solution of our optimization approach coincides with the result, obtained by selecting all features, here denoted by "ALL".  

From this analysis, we conclude, that a high value for the AUROC can be obtained when using the ROC-dependency measure between the features and the  correlation dependency measure between each feature and the target vector, i.e., the combination (ROC, Correl). Furthermore, we find that a rather high value of $\phi \approx 0.95$ should be used. 

The breast cancer data set leads to  somewhat different results, which shows the well-known dependency of machine learning results on individual data sets. In Fig.~\ref{fig:bcbestphi}, the overall $\phi$-dependence of the AUROC for different dependency measures is shown. Here, we find the highest value of the AUROC for the combination (MI, MI), but only for a narrow band of $\phi$-values around 0.75. The next best combination is (ROC, Correl), as in the case of lending club data.

Fig. \ref{fig:bcbestphi} shows the best and the second best constellation - compared to the brute-force, RFE, and LASSO results. Again, with all selection methods investigated the best value of AUROC can not be achieved. However, with the optimization approach (and a suitable choice of parameters) we can beat RFE as well as LASSO.  

To summarize our results so far:
\begin{itemize}

\item For both data sets it was not possible to determine the best subset of features (as obtained by brute-force) with the classical selection methods (RFE and LASSO) considered.

\item Using feature selection via optimization may be superior over these classical methods but the results are highly dependent on the dependency measures used and the hyper-parameter $\phi$.

\item For both data sets we found better results for higher values of $\phi$ which indicates that the dependence of the features with the target seems to be more important than the intra-feature dependence.

\end{itemize}

\section{Feature Selection within the gate-based quantum computer framework}
\subsection{Quantum Algorithms for Optimization}

The optimization problem in equation (\ref{QUBO}) is the starting point for transferring the problem to a quantum computer in the framework of the gate-based approach. The binary variables $z_i$ are converted to operators with eigenvalues 
$\{1;-1\}$ and eigenstates $\vert 0 \rangle $ and $\vert 1 \rangle$. In this way, by applying the transformation $z_i=(1+s_i)/2$ our problem is mapped to finding the ground state of the Ising-like model

\begin{equation}\label{Ising}
	{\cal H} = \sum_{i,j} J_{i,j} s_i s_j~,
\end{equation}

where the coupling matrix $J_{i,j}$ is derived from the above introduced matrix $Q$. 

For solving these types of problems on a gate-based computer several algorithms have been proposed. Here, we mainly use the QAOA algorithm, which has been suggested by Farhi et al.~\cite{Farhi2014}.

The key idea of QAOA is to generate the following quantum state $\vert\psi_{\vec{\gamma},\vec{\beta}}\rangle$ depending on the parameters $\vec{\gamma}=(\gamma_1,\dots,\gamma_p)$ and $\vec{\beta}=(\beta_1,\dots,\beta_p)$ (with a number of iterations $p$):
\begin{equation}
\vert\psi_{\vec{\gamma},\vec{\beta}}\rangle_M=\hat{U}_M(\beta_p)e^{-i \gamma_p \hat{F}}\dots \hat{U}_M(\beta_2)e^{-i \gamma_2 \hat{F}}\hat{U}_M(\beta_1)
e^{-i \gamma_1 \hat{F}}
\vert\psi_0\rangle_M~,\label{eq:QAOA}
\end{equation}
where the initial state $\vert\psi_0\rangle_M$ and the operator $\hat{U}_M(\beta)$ depend on the choice of the mixer $M$.
Here, we use the so-called standard mixer with the Pauli X-matrix
\begin{equation}
\hat{U}_{\rm standard}(\beta)=e^{i\beta \sum_{i=1}^n \hat{X}_i}~.
\end{equation}

After applying the parameterized gates, all qubits are measured with respect to the standard basis which leads to a classical bit-string of zeros and ones. Each bit-string corresponds to a different state and appears with a certain probability. There are $2^n$ different states. 

These bit-strings of several thousand "shots" (we use 8192 shots) are plugged into the objective function eq. (\ref{QUBO_0}) to calculate the expectation value. Then, a classical optimizer is used to update the parameters $\vec{\gamma}$, $\vec{\beta}$, and in order to minimize 
the expectation value and thus the objective function. In other words, to find the ground state of its converted problem Hamiltonian.
A thorough benchmarking study of the QAOA approach, including a description of the solution method for QAOA, is described in great detail by Brandhofer et al.~\cite{Brandhofer2022}.

As a measure to gauge the performance of this optimization procedure we use the approximation ratio:
\begin{equation}\label{Approx.Ratio}
r(z_1,\dots,z_n)= \frac{h(z_1,\dots,z_n)-h_{\rm max}}{h_{\rm min}-h_{\rm max}}~.
\end{equation}
Here $h_{\rm max}$ and $h_{\rm min}$ denote the worst and the best bit-string solution of the problem. If the trained quantum algorithm would always produce the optimal solution with probability 1, the approximation ratio would be $1$. 

To compare the quality of the results, in addition, we use the VQE algorithm \cite{VQE_IBM} with the ansatz of real-amplitudes \cite{RealAmplitudes_IBM}. Basically, the ansatz of real-amplitudes consists of one parametrized R(y)-rotation per qubit and entangling neighboring qubits with a CNOT-gate. Repeating this procedure several times (= number of layers) increases the depth  of the circuit and thereby the complexity. 

All quantum-based calculations in this paper have been performed with IBM's Qiskit framework \cite{Qiskit}. Especially, we use out-of-the-box versions of the QAOA- and VQE-algorithm \cite{QAOA_IBM}, the state vector- and QASM-simulator, as well as different physical quantum backends. The formulation of the optimization problem with IBM's Qiskit \cite{Qiskit} is straightforward.
Within this framework, different solution methods are available for comparison:

\begin{itemize}

\item Solution with IBM's classical optimizer CPLEX \cite{cplex2009v12}.

\item Small problems can be solved by a direct diagonalization of the problem's Hamiltonian.

\item Using the state vector simulator allows to perform the quantum mechanical calculations directly without destroying the quantum state. Again, this is only feasible for small problem sizes.

\item Using a quantum simulator that mimics real hardware, where qubits are prepared, transformed by several gates, and then measured. From the measurements, the solution to the problem can be derived. Here, we use the noise-free QASM simulator within IBM's framework which is available in the IBM cloud for sizes up to 32 qubits.

\item Using a physical quantum computer in the cloud to perform the calculation on a real device. 

\end{itemize}

For small problem sizes, as the one considered up to now, all solution strategies described can be used to benchmark the results. However, the bigger the problem, the fewer are the range of usable strategies. From a problem size of approx. 50 qubits, the exact diagonalization, and the state vector and quantum simulation methods are not available anymore and the only remaining possibility is using CPLEX (running classical optimization routines under the hood) or using real quantum hardware. 

\subsection{Results on quantum simulators for different dependency measures and $\phi$-values}

\subsubsection{Lending Club Data}
As a first step, we compare the results shown in Table~\ref{tab:resTuplesLending} for the case of the lending club data set with different tuples $(\rho_{Feature}^X, \rho_{Target}^Y)$.
Interestingly, for all the applied solution methods and algorithms (CPLEX, exact diagonalization, state vector-simulator, QASM-simulator, QAOA- and VQE-algorithm) the same optimization value (given $(\rho_{Feature}^X$ and $\rho_{Target}^Y)$) is obtained. Furthermore, the calculated accuracy is for all dependency measures equal to $0.830$. However, as shown in the table above, the number of selected features and the area under curve (AUROC) differ.  

\begin{table}
    \centering
    \caption{Lending club data: performance of the optimization approaches when selecting features for the logistic regression model with different dependency measures.}
    \label{tab:resTuplesLending}
\begin{tabular}{|c|c|c|c|c|}
\hline	
$\rho_{Feature}^X$ &  $\rho_{Target}^Y$ & Optimization value & AUROC & Number of features\\
\hline
$X=Correl$  & $Y=Correl$ & -0.3542 & 0.6635 & 5\\
\hline
$X=Correl$ & $Y=MI$ & -0,0617 & 0.6405 & 3  \\
\hline
$X=Correl$ & $Y=ROC$ & -3.7803&  {\bf 0.6651} & 8\\
\hline
$X=Correl$ & $Y=ANOVA$ & -87.8750 &  0.6634 & 7 \\
\hline 
$X=MI$  & $Y=Correl$ & -0.4117 & 0.6639 & 6\\
\hline
$X=MI$ & $Y=MI$ &-0.0445  & 0.5790 & 1\\
\hline
$X=MI$ & $Y=ROC$ & -4.0131 &  {\bf 0.6651} & 8\\
\hline
$X=MI$ & $Y=ANOVA$ & -88.0025  &  0.6634 &7 \\
\hline 
\end{tabular}
\label{Approx.RatioQ}
\end{table}

\begin{figure}[!h]
\centering
\includegraphics[width=15cm]
{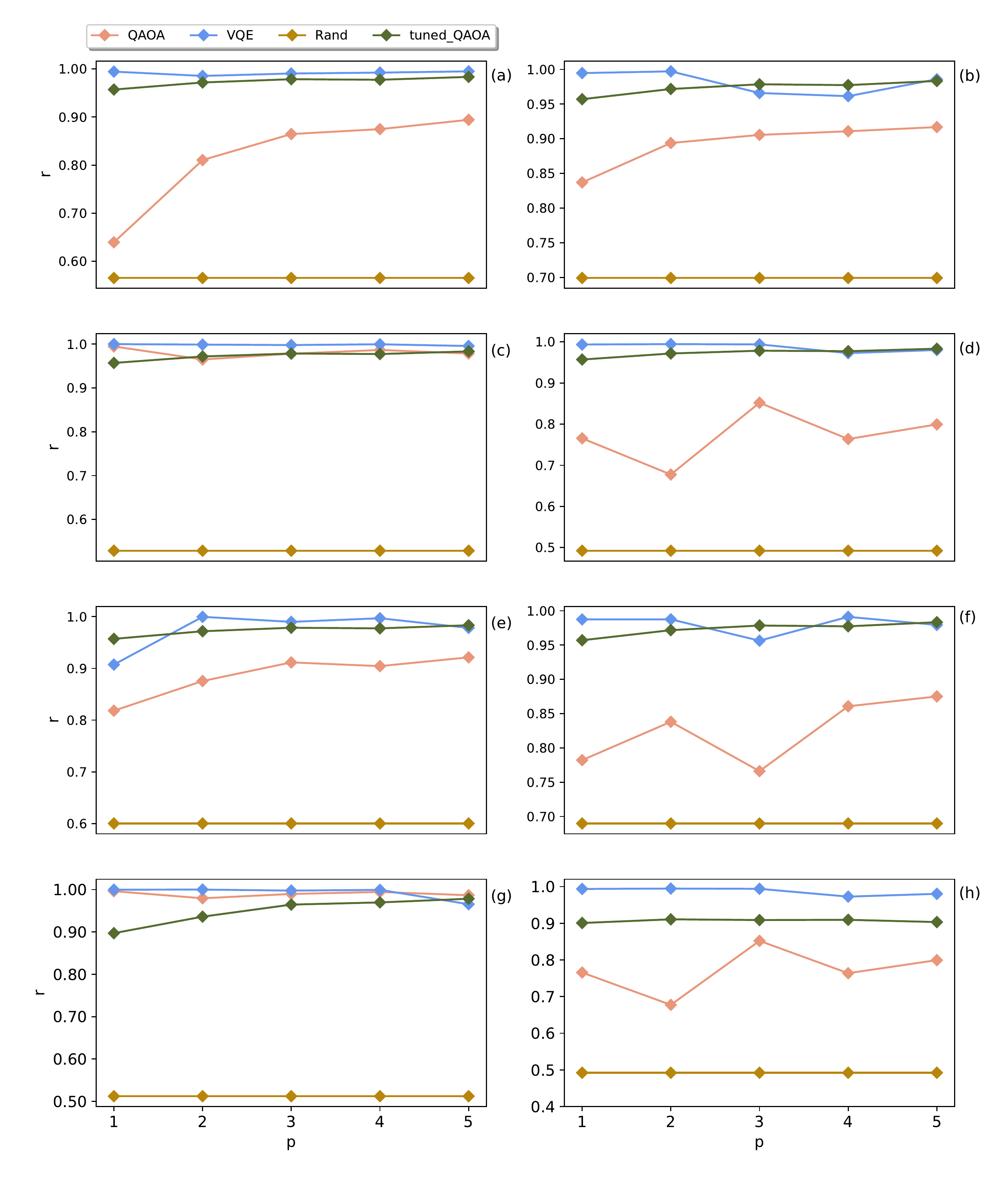}
\caption[Lending Club data: Approximation ratio obtained with VQE, QAOA and tuned QAOA for different values of p, $\phi=0.95$ and 
for the following dependency measures: (a) $(\rho_{Feature}^{Correl},\rho_{Target}^{Correl})$, (b) $(\rho_{Feature}^{Correl},\rho_{Target}^{MI})$, (c) $(\rho_{Feature}^{Correl},\rho_{Target}^{ROC})$, (d) $(\rho_{Feature}^{Correl},\rho_{Target}^{ANOVA})$, (e) $(\rho_{Feature}^{MI},\rho_{Target}^{Correl})$, (f) $(\rho_{Feature}^{MI},\rho_{Target}^{MI})$, (g) $(\rho_{Feature}^{MI},\rho_{Target}^{ANOVA})$, (h) $(\rho_{Feature}^{MI},\rho_{Target}^{ROC})$
]
{Lending Club data: Approximation ratio obtained with VQE, QAOA and tuned QAOA for different values of p, $\phi=0.95$ and 
for the following dependency measures: (a) $(\rho_{Feature}^{Correl},\rho_{Target}^{Correl})$, (b) $(\rho_{Feature}^{Correl},\rho_{Target}^{MI})$, (c) $(\rho_{Feature}^{Correl},\rho_{Target}^{ROC})$, (d) $(\rho_{Feature}^{Correl},\rho_{Target}^{ANOVA})$, (e) $(\rho_{Feature}^{MI},\rho_{Target}^{Correl})$, (f) $(\rho_{Feature}^{MI},\rho_{Target}^{MI})$, (g) $(\rho_{Feature}^{MI},\rho_{Target}^{ANOVA})$, (h) $(\rho_{Feature}^{MI},\rho_{Target}^{ROC})$}
\label{fig:AR_00}
\end{figure}

In Fig. \ref{fig:AR_00}, we show the dependency of the approximation ratio on the number of layers - for QAOA as well as for VQE. For comparison, we also show the approximation ratio of random search. This value is obtained by drawing 1000 uniform random bit-strings and calculating the mean approximation ratio. 

The approximation ratios obtained with the VQE algorithm are in almost all cases close to 1 and are in general higher than the results obtained with QAOA.  Further, the QAOA results are distinct: In some cases, the approximation ratios are - independent of $p$ - near 1, in other cases, they increase  with increasing $p$. This supports the conjecture raised in \cite{Brandhofer2022} that there are easy and not-so-easy parameter combinations. From a practical point of view, it would be desirable to have on the one hand a QUBO matrix that leads to a high value of AUROC (see table \ref{Approx.RatioQ}) and on the other hand, a QUBO matrix that is easy to handle for the calculation, i.e. it leads to high approximation ratios with only a few iterations $p$. Taking this into account, $(\rho_{Feature}^{Correl},\rho_{Target}^{ROC})$ seems to be a good choice for this problem. 

\begin{figure}[tbh]
	\centering
	\includegraphics[scale=0.85]{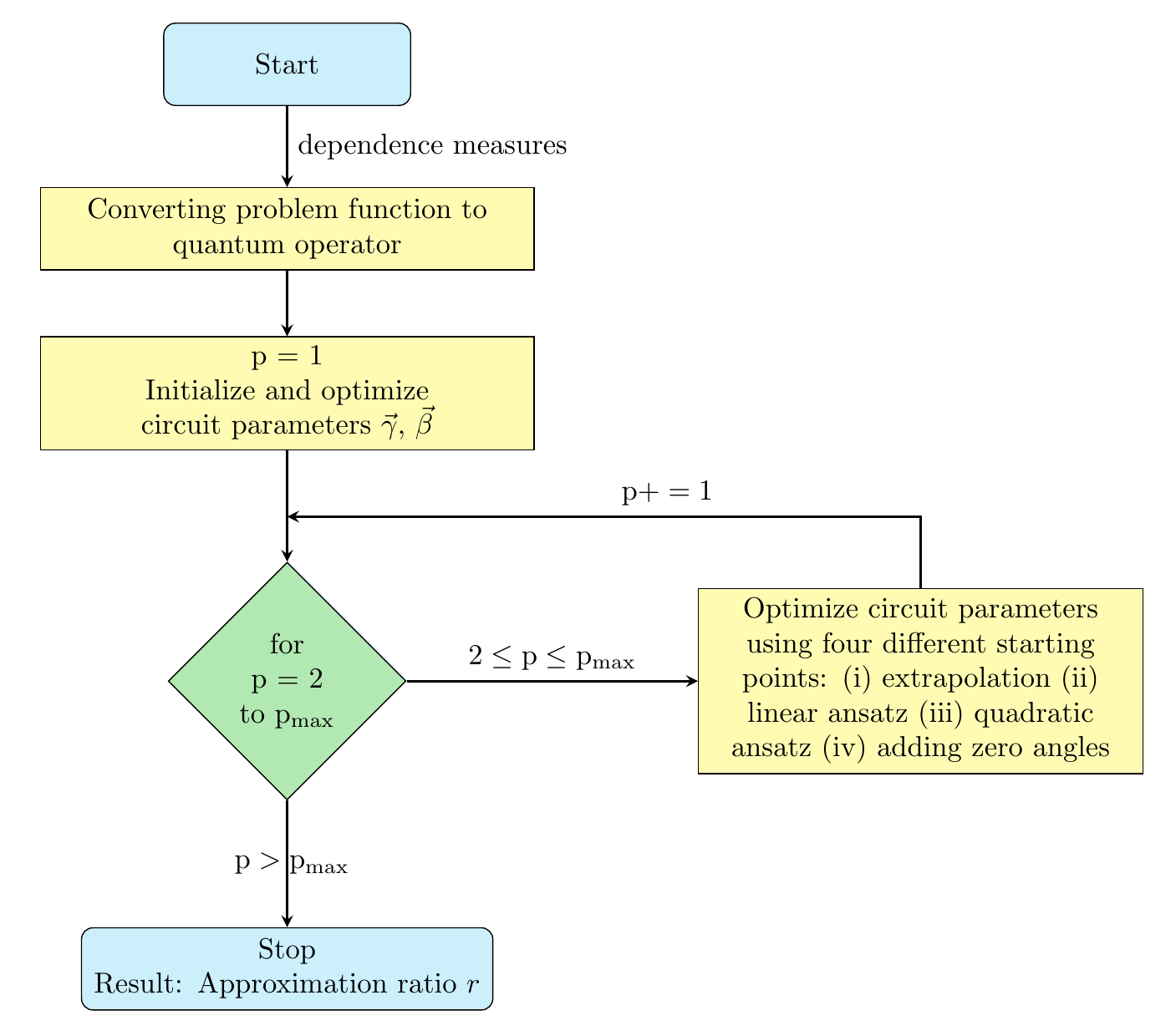}
	\caption{\small Flowchart of the QAOA routine for solving the Feature Selection problem, whereby the particularity is to find suitable candidates for initial values of the circuit parameters using four heuristic methods described in \cite{Brandhofer2022}}
	\label{fig:flowchart}
\end{figure}

\newpage
In order to improve the performance of QAOA, we make use of four heuristic methods ((i) extrapolation, (ii) linear ansatz, (iii) quadratic ansatz, (iv) adding zero angles) to choose suitable candidates for initial values of the circuit parameters $\gamma$ and $\beta$ for increasing QAOA layers $p$ as proposed in \cite{Brandhofer2022}. By repeating the optimization runs with different initial values one can avoid getting stuck in a local minimum. Further, it has been shown that the required time scales exponentially faster to achieve a similar performance when using the observed patterns of the optimal parameters instead of using arbitrary starting values \cite{zhou20}.
The QAOA routine as proposed in \cite{Brandhofer2022} and which is used here to solve the feature selection problem is depicted in Fig.~\ref{fig:flowchart}. 

\begin{figure}[!h]
\centering
\includegraphics[width=15cm]{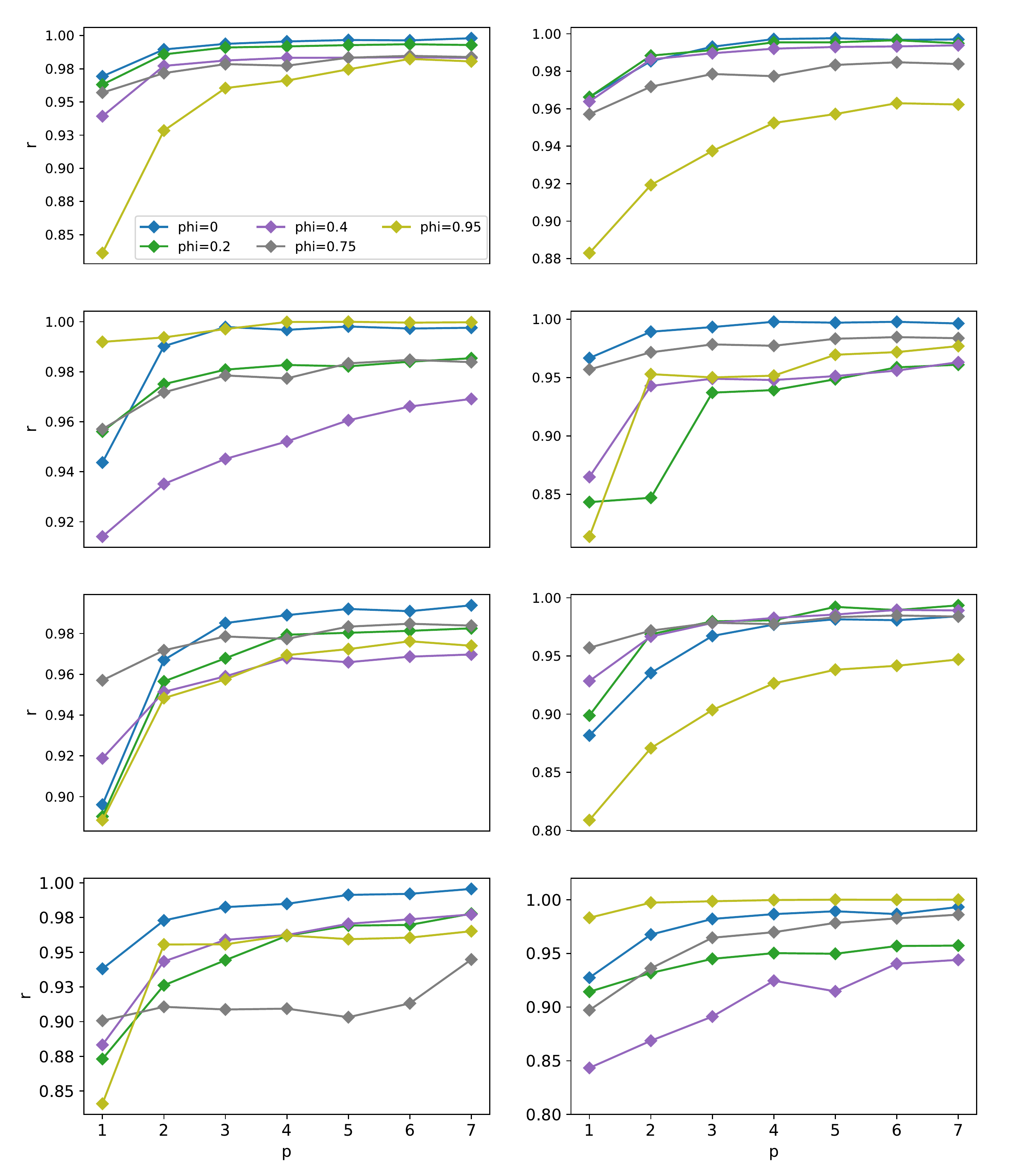}
\caption[Lending Club data: Approximation ratio for different values of $\phi$ shown for different values of QAOA layers p evaluated for the following dependency measures: (a) $(\rho_{Feature}^{Correl},\rho_{Target}^{Correl})$, (b) $(\rho_{Feature}^{Correl},\rho_{Target}^{MI})$, (c) $(\rho_{Feature}^{Correl},\rho_{Target}^{ROC})$, (d) $(\rho_{Feature}^{Correl},\rho_{Target}^{ANOVA})$, (e) $(\rho_{Feature}^{MI},\rho_{Target}^{Correl})$, (f) $(\rho_{Feature}^{MI},\rho_{Target}^{MI})$, (g) $(\rho_{Feature}^{MI},\rho_{Target}^{ANOVA})$, (h) $(\rho_{Feature}^{MI},\rho_{Target}^{ROC})$ ]{Landing Club data: Approximation ratio for different values of $\phi$ shown for different values of QAOA layers p evaluated for the following dependency measures: (a) $(\rho_{Feature}^{Correl},\rho_{Target}^{Correl})$, (b) $(\rho_{Feature}^{Correl},\rho_{Target}^{MI})$, (c) $(\rho_{Feature}^{Correl},\rho_{Target}^{ROC})$, (d) $(\rho_{Feature}^{Correl},\rho_{Target}^{ANOVA})$, (e) $(\rho_{Feature}^{MI},\rho_{Target}^{Correl})$, (f) $(\rho_{Feature}^{MI},\rho_{Target}^{MI})$, (g) $(\rho_{Feature}^{MI},\rho_{Target}^{ANOVA})$, (h) $(\rho_{Feature}^{MI},\rho_{Target}^{ROC})$}
\label{fig:AR_1}
\end{figure}

In Fig.~\ref{fig:AR_1}, the approximation ratios for different values of $\phi$ for the lending club data are given as a function of the number of QAOA layers $p$.
For this data set, it is noticeable that the majority of the dependency measures (Fig.~\ref{fig:AR_1} (a), (b), (d), (e), (g)) a value of $\phi=0$ lead to the fastest convergence towards the optimal solution, whereas higher values of $\phi$ (e.g. $0.75 \leq \phi < 0.95$) lead to a slower convergence.
In other words, we find that the dependence between the features is more significant than the dependence between features and the target. For the other dependency measures (Fig.~\ref{fig:AR_1} (c), (f), (h)), we find that large values of $\phi$ (e.g.
$0.75 \leq \phi < 1$) perform better than values of $\phi \leq 0.5$. Consequently, the weight for the two conditions (intra-feature dependence and dependence between features and target) depends on the choice of the dependency measure and it is not useful to be set globally in order to achieve the best possible performance - it is rather suggested to set a local value for the weighting parameter depending on the individual dependency measure.

Lastly in this section, we present the results obtained using a physical quantum backend of IBM for a fixed parameter setup and with the ROC-dependence between the features and the target, and the linear correlation between all features. Furthermore, we set $\phi=0.95$.

Using the lending club data, we obtain the same optimal value for the objective function, $obj=-3.7803$, with the classical optimizer, the direct diagonalization of the Hamiltonian, the state vector simulator, and with the quantum simulator. Further, we also obtain the same value for the AUROC, i.e. $AUROC=0.6651$, c.f. (\ref{Approx.RatioQ}). This corresponds to the selection of all features.

When using a physical quantum backend (here: the IBM system Montreal) with $p=1$, we obtain the same objective function value of $-3.7803$. However, the approximation ratio drops to 0.59. With $p=5$, the approximation ratio increases to 0.66. These values are to be compared with 0.52, the approximation ratio obtained by randomly selecting bit-strings.
While these results reflect the errors of existing real-world QC backends, they show nevertheless the expected dependence on $p$.

\newpage

\subsubsection{Breast Cancer Data}
We now repeat the calculation from the last subsection for the breast cancer data. The results for the different tuples $(\rho_{Feature}^X, \rho_{Target}^Y)$ are reported in Table~\ref{Approx.RatioQ_Cancer}.
\begin{table}
    \centering
    \caption{Breast cancer data: performance of the optimization approach when selecting features for the logistic regression model with different dependency measures.}    
 \label{Approx.RatioQ_Cancer}
\begin{tabular}{|c|c|c|c|c|c|}
\hline	
$\rho_{Feature}^X$ &  $\rho_{Target}^Y$ & Optimization value & AUC & ACC & Number of features \\
	\hline
	$X=Correl$  & $Y=Correl$ &-0.91611 & 0.9806 & 0.9209 & 3\\
	\hline
	$X=Correl$ & $Y=MI$ &  -0.3300 & 0.9644 & 0.8910  & 1  \\
	\hline
	$X=Correl$ & $Y=ROC$ & -1.7031 &  0.9822  & 0.9192 & 4 \\
	\hline
	$X=Correl$ & $Y=ANOVA$ & -2913.6696 &  0.9839 & {\bf 0.9315} &9 \\
	\hline 
	$X=MI$  & $Y=Correl$ & -1.438 & 0.9796 & 0.9262 & 5\\
	\hline
	$X=MI$ & $Y=MI$ &-0.4139  & {\bf 0.9859} & 0.9297 &3\\
	\hline
	$X=MI$ & $Y=ROC$ & -2.6366 &  0.9828 & 0.9279 & 6\\
	\hline
	$X=MI$ & $Y=ANOVA$ & -2914.5083  &  0.9839 & {\bf 0.9315} & 9\\
	\hline 
\end{tabular}
\end{table}
While the optimization values, given $(\rho_{Feature}^X, \rho_{Target}^Y)$) coincides in general for CPLEX, exact diagonalization, and QAOA (state vector-simulator and  QASM-simulator), we detect that VQE is not always able to find the same optimization value. Interestingly, this behavior depends on the number of layers. I.e. in the case $(\rho_{Feature}^{Correl}, \rho_{Target}^{ROC})$) it occurs that only for $p \leq 3 $ is the VQE algorithm able to find the optimal value. 

Here, in contrast to the lending club data, we see that the accuracy depends on the dependency measure and the highest value for the accuracy can not be obtained together with the highest value of the AUROC.

\begin{figure}[!h]
\centering
\includegraphics[width=15cm]{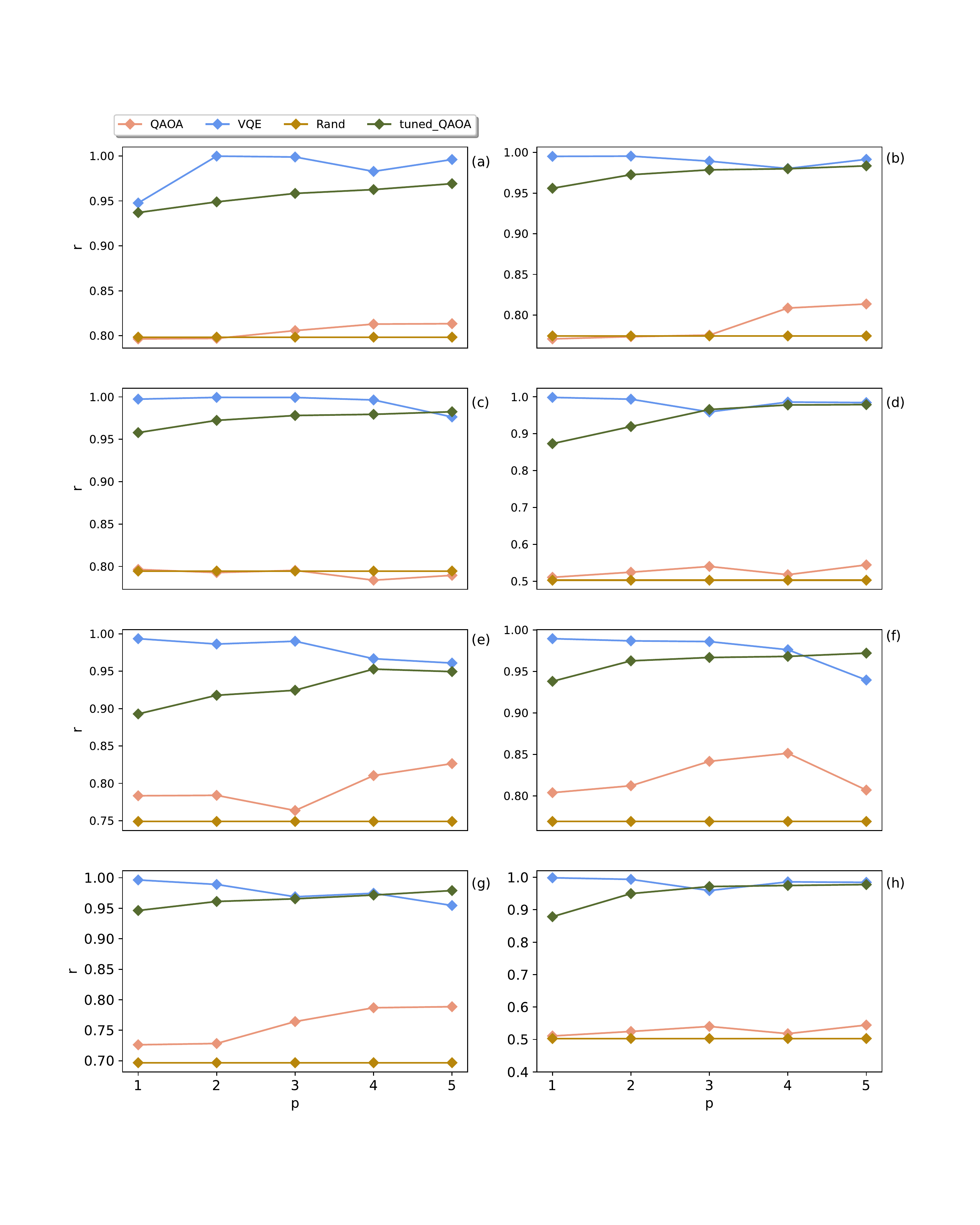}
\caption[Breast cancer data: Approximation ratio obtained with VQE, QAOA and tuned QAOA for different values of p, $\phi=0.75$ and 
for the following dependency measures: (a) $(\rho_{Feature}^{Correl},\rho_{Target}^{Correl})$, (b) $(\rho_{Feature}^{Correl},\rho_{Target}^{MI})$, (c) $(\rho_{Feature}^{Correl},\rho_{Target}^{ROC})$, (d) $(\rho_{Feature}^{Correl},\rho_{Target}^{ANOVA})$, (e) $(\rho_{Feature}^{MI},\rho_{Target}^{Correl})$, (f) $(\rho_{Feature}^{MI},\rho_{Target}^{MI})$, (g) $(\rho_{Feature}^{MI},\rho_{Target}^{ANOVA})$, (h) $(\rho_{Feature}^{MI},\rho_{Target}^{ROC})$
]
{Breast cancer data: Approximation ratio obtained with VQE, QAOA, and tuned QAOA for different values of p, $\phi=0.75$ and 
for the following dependency measures: (a) $(\rho_{Feature}^{Correl},\rho_{Target}^{Correl})$, (b) $(\rho_{Feature}^{Correl},\rho_{Target}^{MI})$, (c) $(\rho_{Feature}^{Correl},\rho_{Target}^{ROC})$, (d) $(\rho_{Feature}^{Correl},\rho_{Target}^{ANOVA})$, (e) $(\rho_{Feature}^{MI},\rho_{Target}^{Correl})$, (f) $(\rho_{Feature}^{MI},\rho_{Target}^{MI})$, (g) $(\rho_{Feature}^{MI},\rho_{Target}^{ANOVA})$, (h) $(\rho_{Feature}^{MI},\rho_{Target}^{ROC})$}
\label{fig:BC_AR_00}
\end{figure}

\begin{figure}[!h]
\centering
\includegraphics[width=15cm]{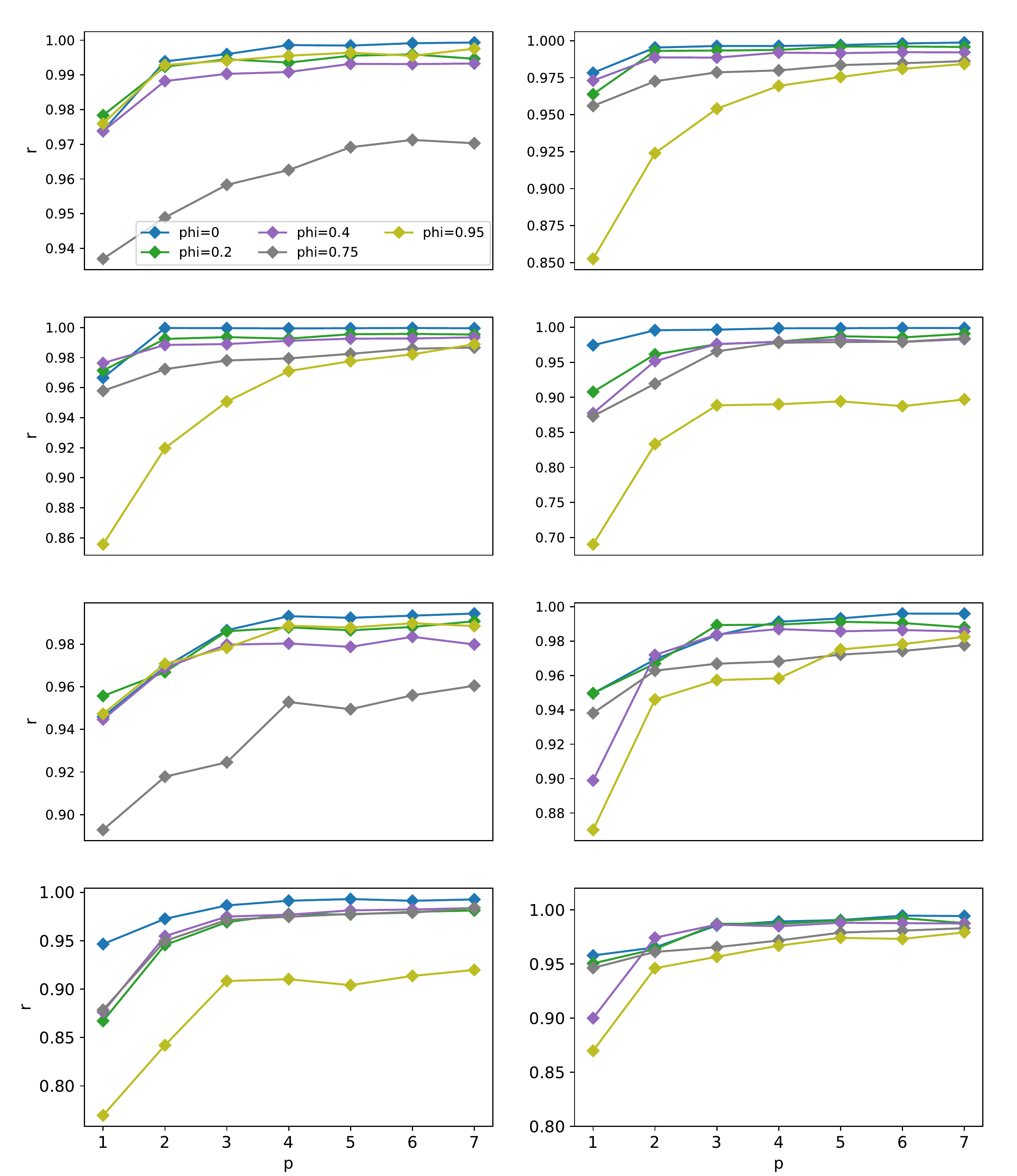}

\caption[Breast cancer data: Approximation ratio for different values of $\phi$ shown for different values of QAOA layers p evaluated for the following dependency measures: (a) $(\rho_{Feature}^{Correl},\rho_{Target}^{Correl})$, (b) $(\rho_{Feature}^{Correl},\rho_{Target}^{MI})$, (c) $(\rho_{Feature}^{Correl},\rho_{Target}^{ROC})$, (d) $(\rho_{Feature}^{Correl},\rho_{Target}^{ANOVA})$, (e) $(\rho_{Feature}^{MI},\rho_{Target}^{Correl})$, (f) $(\rho_{Feature}^{MI},\rho_{Target}^{MI})$, (g) $(\rho_{Feature}^{MI},\rho_{Target}^{ANOVA})$, (h) $(\rho_{Feature}^{MI},\rho_{Target}^{ROC})$ ]{Breast cancer data: Approximation ratio for different values of $\phi$ shown for different values of QAOA layers p evaluated for the following dependency measures: (a) $(\rho_{Feature}^{Correl},\rho_{Target}^{Correl})$, (b) $(\rho_{Feature}^{Correl},\rho_{Target}^{MI})$, (c) $(\rho_{Feature}^{Correl},\rho_{Target}^{ROC})$, (d) $(\rho_{Feature}^{Correl},\rho_{Target}^{ANOVA})$, (e) $(\rho_{Feature}^{MI},\rho_{Target}^{Correl})$, (f) $(\rho_{Feature}^{MI},\rho_{Target}^{MI})$, (g) $(\rho_{Feature}^{MI},\rho_{Target}^{ANOVA})$, (h) $(\rho_{Feature}^{MI},\rho_{Target}^{ROC})$}
\label{fig:AR_2}
\end{figure}

For comparison, the approximation ratios for the breast cancer data are shown in Fig. 
\ref{fig:AR_2}. Again, different values of $\phi$ are used and we show the dependence on the number of QAOA layers. For the dependency measures $(\rho_{\text{Feature}}^{\text{Correl}}, \rho_{\text{Target}}^{\text{Correl}})$ and $(\rho_{\text{Feature}}^{\text{MI}}, \rho_{\text{Target}}^{\text{Correl}})$ we find that setting the parameter $\phi$ to $0.75$ leads to the slowest convergence towards the optimal solution, whereas for the other measures a value of $\phi=0.95$ is showing this behavior. The best convergence is achieved in all cases for an intra-feature setting ($\phi=0$) that opens up the question of what influence the condition between feature and target should have, while the results show that this condition does not play a role in order to achieve the best-optimized solution. Compared to the lending club data set, we find that the approximation ratios for the different values of $\phi$ show a more continuous behavior in terms of the choice of the dependency measure.

As a final result, we present the solution for a fixed parameter combination on a physical quantum backend of IBM. The results are obtained with the mutual information dependence between the features and also to the target. Furthermore, we fix $\phi =0.75$.
Due to limitations in the use of physical quantum backends, we restrict ourselves to two cases and solved the problem with the QAOA algorithms with different values of $p$. 

1) $(\rho_{Feature}^{Correl}, \rho_{Target}^{Correl})$:  
As backends, we use the IBM systems Toronto ($p=1,5$), Auckland ($p=2,3$), and Hanoi ($p=4$). We are aware of the fact that results from different machines are not directly comparable. However, due to the availability of system resources, it was not possible to perform all calculations on the same hardware. As approximation ratios  we get: $p=1: 0.8981; p=2: 0.8734; p=3: 0.8917; p=4: 0.8562; p=5: 0.8706$. These values have to be compared  with the random approximation ratio of $0.7984$, which is significantly lower for all values of $p$. The approximation ratio obtained on the physical backends is in the order of magnitude as obtained with the QASM simulator.

2) $(\rho_{Feature}^{Correl}, \rho_{Target}^{MI})$: Again, we obtained the same value for the objective function and the same features selected as reported in Table~\ref{Approx.RatioQ_Cancer}. As backends, we use the IBMQ systems Toronto ($p=1$) and Geneva ($p=2,3,4,5$).  As approximation ratios  we get: $p=1: 0.7897; p=2: 0.8348; p=3: 0.7311; p=4: 0.7620; p=5: 0.7491$. These values have to be compared  with the random approximation ratio of $0.7743$. The approximation ratio obtained on the physical backends is in the order of magnitude obtained with the QASM simulator.

In the next section, we will explore the results obtained in the context of a larger problem set. 

\section{Extension to 27 features aka qubits}

The results obtained so far are encouraging and we extend the calculation to a larger problem. Here, we use a subset of 27 features of the data considered in \cite{Milne2017}.

For this problem size, using a brute-force approach to determine the best selection of features is not possible anymore. In addition, the state vector-simulator and the exact diagonalization of the Hamiltonian are infeasible. Therefore, we compare the following approaches before applying feature selection via optimization:

\begin{itemize}
	
	\item Calculate the AUROC with all features,
	\item Select the most relevant features via RFE and calculate the AUROC,
	\item Select the most relevant features via  LASSO and calculate the AUROC.	
\end{itemize}

\begin{table}
    \centering
    \caption{Model performance for the data with 27 features.}    
 \label{tab:model27}
\begin{tabular}{|c|c|c|c|}
	\hline
	& AUROC & Accuracy & Number of features \\
	\hline
	All features & 0.7913 & 0.764 & 27\\
	\hline
	Selection with RFE & 0.7904 & 0.757 & 17 \\
	\hline
	Selection with LASSO & 0.7768 &  0.732 &  17\\
	\hline
\end{tabular}
\end{table}

The results are shown in Table~\ref{tab:model27}.
Again, these values serve as benchmarks and have to be compared with the results obtained according to the QUBO scheme.
We translate the feature selection problem to an optimization problem, using correlation as the relevant dependence measure between the features and between features and target and we fix $\phi=0.9$. 

Then, the following strategies are used to solve the optimization problem:
	
\begin{itemize}
	
	\item Select the features with a greedy algorithm, starting from a random configuration and then flipping the qubits to improve the objective function
	
	\item Solve the optimization problem with a classical optimizer (here: IBM's CPLEX).
	
	\item Solve the problem with a quantum simulator, available in the IBM cloud with using the QAOA algorithm. 
	
	\item Solve the optimization problem using the QAOA algorithm with a 27-qubit real quantum device.

\end{itemize}

In contrast to the last chapter, where we used a tuned QAOA algorithm, we use here IBM's standard implementation: First, the advantages of the tuned algorithm are only observable at higher values of p.  Since we are now looking at 27 qubits, a useful computation for higher values of p is no longer possible, so we can no longer take advantage of this variant and choose to use IBM's standard implementation instead.
Second, when using the IBM algorithms, the cloud infrastructure and the interplay between classical and quantum algorithms via the so-called Runtime can be optimally exploited. 

\begin{table}
    \centering
    \caption{Results for the data with 27 features, with classical optimization approaches.}    
 \label{tab:model27optim}
\begin{tabular}{|c|c|c|c|c|}
	\hline
	& Objective & AUROC & Accuracy & Number of features \\
	\hline
	Greedy optimization & -0.7232 & 0.7469 & 0.709 & 8\\
	\hline
	IBM's CPLEX & -0.7232 & 0.7469  & 0.709 & 8 \\
	\hline
\end{tabular}
\end{table}

The results using the greedy algorithm and IBM’s CPLEX optimizer are presented in Table~\ref{tab:model27optim}.
The calculation with the real quantum device shown below should be considered a proof of concept. For the calculation of the approximation ratio, we use the optimal CPLEX result as ground truth.

We use different US devices available in the IBM cloud - Auckland and  Kolkata  as well as the German system Ehningen. Results are shown in Table~\ref{tab:model27optimQ}.
\begin{table}
    \centering
    \caption{Results for the data with 27 features, with QC optimization approaches.}  
 \label{tab:model27optimQ}
\begin{tabular}{|c|c|c|c|c|}
	\hline
	& Objective & AUROC & Accuracy & Number of features \\
	\hline
		
	QUASM-Simulator, p=1& -0.6821 & 0.7412 & 0.731 & 7 \\
	\hline
	QUASM-Simulator, p=2&  -0.6974 & 0.718 &  0.707  & 7 \\
	\hline
	QUASM-Simulator, p=3 & -0.70421  & 0.7435 & 0.707& 7 \\
	\hline
	Real Device (Auckland), p=1 & -0.6097 & 0.7323  & 0.722 & 5 \\
	\hline
	Real Device (Auckland), p=2 & -0.6218 & 0.7444 & 0.709 & 8 \\
	\hline
	Real Device (Auckland), p=3 & -0.6348 &  0.7469 & 0.7225 & 6 \\
	\hline
	Real Device (Kolkata), p=4 & -0.6199 &  0.7482 & 0.72 & 9\\
	\hline 
	Real Device (Ehningen), p=1 & -0.6073 & 0.7374 & 0.702 & 7 \\
	\hline
	Real Device (Ehningen), p=2 & -0.5982 & 0.7304 & 0.695 & 7 \\
	\hline
	Real Device (Ehningen), p=3 & -0.3543 &  0.7452 & 0.712 & 7 \\
	\hline
\end{tabular}
\end{table}

Again, we also calculate the random approximation ratio, calculated with a sample of 1000 bit-strings. 

Due to the fact that now we do not necessarily get the same optimal objective  value compared with CPLEX, we use two different prescriptions: 

\begin{itemize}
	
	\item Approximation ratio rel. CPLEX: We take the best and the worst objective value out of CPLEX as ground truth and calculate the approximation ratios relative to these values. 
	
	\item Approximation ratio rel. Sim.: From the outcome of the simulations (either random sampling as a baseline or the quantum simulations) we take the best and the worst value of the objective function and use those in the formula for the approximation ratio. This number describes the spread of the different simulation results.
\end{itemize}
The respective values are shown in Table~\ref{tab:model27optimQAR}.
\begin{table}
    \centering
    \caption{Approximation ratios for the data with 27 features, with QC optimization approaches.}     
    \label{tab:model27optimQAR}
\begin{tabular}{|c|c|c|}
	\hline
	& Approx. ratio rel. CPLEX & Approx. ratio rel. Sim. \\
	\hline
	Random Sampling & 0.7314 &  0.6899\\
	\hline
	Real Device (Auckland), p=1 & 0.8351&  0.7086 \\
	\hline
	Real Device (Auckland), p=2 & 0.8189 & 0.7466 \\
	\hline
	Real Device (Auckland), p=3 & 0.8274 & 0.7159   \\
	\hline
    Real Device (Kolkata), p=4 & 0.8271 &  0.7272 \\
	\hline
	Real Device (Ehningen), p=1 &  0.8356 & 0.7185\\
	\hline
	Real Device (Ehningen), p=2 & 0.8734 &  0.6941\\
	\hline
	Real Device (Ehningen), p=3 & 0.8541 & 0.7299 \\
	\hline
\end{tabular}
\end{table}
The results obtained with the physical quantum devices are surprisingly good, considering the existing error rates. The minimum objective function as determined with CPLEX is almost achieved in most cases. The approximation ratios are well above the random sampling result and we see a slight improvement when increasing the value of $p$. 

\subsection{Comparison to classical/metaheuristic optimization algorithms}

In order to gain more insight into the details of this optimization problem, we perform another experiment. 
Here, we are interested in a comparison between classical, metaheuristic algorithms, such as evolutionary algorithms (EA) or estimation of distribution algorithms (EDA)~\cite{Eiben2003,Blum2003,Hauschild2011,Boussaid2013}.

One core element of such algorithms are stochastic components, e.g., the random mutation of solutions in an EA or the sampling from a distribution in an EDA.
Similarly to non-quantum metaheuristics, QAOA also has random components (potentially as part of the classical update step, as well as stochastic elements in the physical quantum circuit).
As such, a comparison to (mostly) deterministic solvers such as CPLEX is less intuitive.
Moreover, globally optimal solutions might not be necessary (and in fact not guaranteed to be found within a limited time frame), and hence it may be of interest to investigate how intermediate (non-optimal) results from a classical stochastic algorithm compare to those of QAOA.

Another interesting aspect of this comparison is that the  metaheuristics we consider are \textit{not} limited to quadratic (or polynomial) objective functions: they are also applicable to arbitrary non-linear optimization problems (of course, with the respective consequences for performance).

\subsection{Classical, stochastic optimization algorithms}
We tested four metaheuristics: A simple evolutionary algorithm (EA), a self-adaptive evolutionary algorithm (SAEA),
a univariate estimation-of-distribution algorithm (UEDA), and a discrete version of the covariance matrix adaption evolution strategy (DCMA).

\subsubsection{Evolutionary algorithm}
Evolutionary algorithms (EA) iteratively generate 
new candidate solutions ("offspring") from existing candidate solutions ("parents") via random variation operators ("mutation" and "recombination"). Well-performing candidates influence candidates in the next iteration. For a more in-depth introduction see, e.g.,~\cite{Eiben2003}.

The simple EA employed in this experiment uses a fixed mutation and recombination rate, considering a population of $\mu$ parents. Initially, candidate solutions are generated by first sampling the number $n_{bit}$ uniformly from ${1,2,...,27}$, then uniformly and randomly generating a bit-string with exactly $n_{bit}$ ones.

In each iteration, new candidate solutions are generated via a mutation operator and a recombination operator.
The following operators are used:
\begin{itemize}
    \item  Mutation: bit-flip; inverting a random number of bits of an existing candidate solution.
    \item  Mutation: block-inversion; inverting a random block (sequence) ${i, i+1, ... j-1,j}$ of bits.
    \item  Mutation: cycle; cyclically shifting bit-values to the right or left.
    \item Recombination: one-point crossover; all bits up to bit $i$ are taken from one parent, the remaining bits from the second parent.
    \item Recombination: two-point crossover; selecting a random sequence of bits ${i, i+1, ... j-1,j}$ from one parent, the remaining bits from the second parent.
    \item Recombination: uniform crossover; selecting bits uniform randomly from both parents.
\end{itemize}

The mutation rate $r_m$ determines how many bits are changed with each mutation operator (i.e., number of bits flipped, size of the inverted block, step size of the cyclical shift).
We will investigate four parameters for tuning: population size $\mu$, mutation operator choice for recombination $o_r$ and mutation $o_m$, and the mutation rate $r_m$.
The remaining parameters of the EA are left at default values.
For performance reasons, an archive of all candidate solutions is not retained. The employed implementation is available via the R-package \texttt{CEGO}~\cite{Zaefferer.CEGO.2.4.2}.

\subsubsection{Self-adaptive evolutionary algorithm}
One issue of the 'simple' EA described above is that it requires setting fixed algorithm parameters. The performance of the EA is rather sensitive to some of these parameters (e.g., population size, operator choices, and mutation rate).
Hence, these parameters may require tuning or else, may lead to unsatisfactory performance.

The self-adaptive EA (SAEA) tries to partially alleviate this issue by attaching several parameters $\theta$ to each solution candidate (i.e., to each individual in a population).
Our SAEA variant follows a similar approach as described for the Mixed Integer Evolution Strategy~\cite{Li2013}.

Subsequently, the parameters $\theta$ are evolved alongside the actual binary decision variables $z$: 
each candidate solution is then composed as $z'=\{z,\theta\}$.
Thus, if some solution candidate is successful (i.e., has a good objective function value), the associated parameters $\theta$ (e.g., the mutation rate) that were used to generate it will be more likely to be used in subsequent iterations as well.

Note, that some algorithm parameters will necessarily be fixed (at least during a single algorithm run), 
and not attached to each candidate solution. For example, we have to specify a set of operators, ranges of parameters, and
learning rates to configure the self-adaptive procedure itself.
The underlying motivation is that the algorithm should be less sensitive to the parameters that control the self-adaptive procedure (e.g., learning rate) when compared to the sensitivity to changes in the parameters controlled by that procedure (e.g., mutation rate).

In detail, the SAEA will self-adapt the following three parameters: the mutation rate $\theta_1=r_m \in [1/n, 1]$,
the mutation operator choice $\theta_2=o_m$ (bit-flip, block-inversion, cycle), and the recombination operator choice $\theta_3=o_r$  (1-point, 2-point, uniform).

The choices of operators for mutation and recombination are themselves also mutated, choosing randomly with probability $p_r$ from the parents.
Similarly, the mutation rate is mutated in each iteration, via $\theta_1^*=\theta_1 e^{\tau \epsilon}$. 
Here, $\tau$ is a learning rate (hyperparameter) and $\epsilon$ is a normal-distributed random sample with zero mean and unit variance. 
Intermediate crossover is used to recombine $\theta_1^*=f(\theta_{1}^{*(1)},\theta_{1}^{*(2)})$ values from parent solutions $\{z'^{(1)}, z'^{(2)}\}$.

Like the simple EA, the SAEA implementation is available via the R-package \texttt{CEGO}~\cite{Zaefferer.CEGO.2.4.2}, and uses the same initialization strategy.

\subsubsection{Univariate estimation-of-distribution algorithm}
The third algorithm we employ is an estimation of distribution algorithm (EDA).
EDAs essentially model the set of candidate solutions with a distribution, iteratively sampling from that distribution to generate new candidates, then updating the distribution parameters based on the evaluated candidates.

Here, we use a separate univariate binary distribution to model each solution bit $z_i$.
Specifically, our univariate EDA (UEDA) is a variant of the population-based incremental learning algorithm (PBIL)~\cite{Baluja94a}:

\begin{itemize}
  \item Each bit $z_i$ is represented by an independent probability $p_{z_i}$.
  \item New candidate solutions are sampled based on these probabilities. We refer to the number of generated candidate solutions (bit-strings) as the population size $\mu$.
  \item Each candidate solution is evaluated with the objective function.
  \item The better 50\% of the candidate solutions are selected.
  \item Based on that selection, new probabilities $p^*_{z_i}$ are estimated (via the mean value for each bit, over all selected candidates).
  \item The probabilities for the next iteration are set via a 'learning' update $p_{z_i}=p_{z_i}  (1-\tau) + p^*_{z_i} \tau$.
  \item To avoid collapsing to $p_{z_i}=0$ or $p_{z_i}=1$, the probabilities are limited to $p_{z_i} \in [\frac{1}{n},1-\frac{1}{n}]$.
\end{itemize}

An implementation of this simple UEDA was written by the authors in R. It uses the same initialization strategy as the EA.

\subsubsection{Discrete covariance matrix adaption evolution strategy (DCMA)}

Finally, we also employ a variant of the covariance matrix adaption evolution strategy (CMAE-ES), which can handle discrete (including binary) variables by introducing a lower bound on the marginal probabilities~\cite{Hamano2022a}. 
In essence, this algorithm can also be viewed as an EDA. But in contrast to the UEDA, correlations between variables $z_i$ are taken into account, as the internal distribution is based on an evolving covariance matrix.

We use the implementation of this margin-based CMA-ES provided by the \texttt{cmaes} python library~\cite{Shibata2022a} and refer to this algorithm as discrete covariance matrix adaption (DCMA).
Besides the population size $\mu$, this algorithm is also impacted by an initial standard deviation $\sigma_\text{init}$, as well as the margin parameter $\alpha_m$.

\subsubsection{Random and Greedy Search}

Finally, we employ two simple algorithms as baselines to compare against: a simple greedy local search with restarts (GRS) and random search (RS). The former is at its core a purely exploitative, local search, while the latter is purely explorative, global search.

The local search procedure of the GRS starts at a random initial solution (determined in the same way as the initial solutions in the EA). Neighboring candidate solutions are produced by bit-flips, in a random order. 
Once a neighbor improves on the current solution,
it immediately becomes the current solution.
If no neighbor improves on the current solution, the procedure is restarted with a new initial solution.  We do not tune this procedure for the sake of simplicity.
Note that we could in principle make this procedure more complex, e.g., by redefining neighboring solutions to mean something else than single bit-flips, by introducing a more clever way to generate new initial solutions for restarts, or by hybridizing it with the EA or similar algorithms.

The random search simply tries out 
new candidate solutions at random, where
each solution is generated in the same way as random initial solutions in the EAs.

\subsection{Experiment}
The whole data set is randomly split into two subsets. Each subset is used to generate a corresponding instance of the QUBO objective function, using correlation as the respective dependency measure. One instance will be used for tuning, another for validating / comparing the (tuned) algorithms.

For the quantum algorithm QAOA, we run the algorithm on real QC devices (mainly Geneva and Montreal) nine times for $p=3$, solving only the validation instance. Similarly, CPLEX is run on the validation instance of the objective function (and will be used to provide a baseline, as it solves this problem exactly).

\subsubsection{Algorithm Tuning}
Tuning is performed on the first of the two instances of our objective function.
The goal is to provide just a very rough pre-configuration
for the meta-heuristics, to avoid detrimental performance
due to a poor algorithm configuration and allow for a reasonably fair comparison between the metaheuristics.

Therefore, the employed tuning method is uniform random search:
For each algorithm, 100 algorithm configurations are sampled
uniformly at random. 
For each algorithm configuration, 10 independent algorithm runs are performed. Each run uses a maximum of 2000 objective function evaluations.
Afterward, the configuration with the best median performance over the 10 runs is returned as the 'tuned' configuration.  Here, 'performance' means the median sum (overall objective function evaluations (OFE)) of the cumulative minimum objective value recorded by the tuning procedure.
We tune two to four `main' parameters of each algorithm, as specified in Table~\ref{tab:tuning} (omitting the more simple baselines, and omitting QAOA due to computing constraints).

\begin{table}
\centering
\caption{Tuning: bounds and results for each parameter. $b_t$ is the budget, i.e., the maximum number of evaluations per run during tuning. Mutation operator codes are 1: bit-flip, 2: block-inversion, 3: cycle. Recombination operator codes are 1: one-point, 2: two-point, 3: uniform. Where values are specified with exponents, random search applies uniformly in the exponents (instead of uniform in terms of actual values).
The column 'performance' lists the median sum of the cumulative minimum objective value recorded by the tuning procedure. In brackets, the standard deviation over all random configurations is given (not the standard deviation of the best performance only).}
\label{tab:tuning}
\begin{tabular}{l l l l | l} 
 algorithm &parameter & range & tuned & performance\\ [0.5ex] 
  & & of values  & value & (sd-all)\\ [0.5ex] 
 \hline
EA & population size  $\mu$ & $[4, 200]$ & 52 & -565 \\
& mutation rate $r_m$ & $[0, 1]$ & 0.042&(73.2)\\
& mutation operator $o_m$ & $\{1, 2, 3\}$ & 1 &\\ 
& recombination operator $o_r$ & $\{1, 2, 3\}$ & 3 & \\
 \hline
SAEA & population size $\mu$ & $[4,200]$ &  14 & -593 \\
& learning rate $\tau$ & $[10^{-4}, 10^0]$& $10^{-1.32}$ &(25.6)\\
& probability $p_r$ & $[0,1]$& 0.30 & \\ 
 \hline
UEDA & population size $\mu$ & $[4, 200]$& 15 & -586 \\
& learning rate $\tau$ & $[0, 1]$ & 0.95 &(66.5) \\ 
 \hline
DCMA & population size $\mu$ & $[4, 200]$& 19 & -576 \\
& initial stand. dev. $\sigma_\text{init}$ & $[10^{-4},10^4]$ & $10^{-2.03}$ &(59.0) \\
& margin $\alpha_m$ & $[(\mu n)^{-1.5}, (\mu n)^{-0.5}]$ & $(\mu n)^{-1.23}$  &\\
\end{tabular}
\end{table}

The tuning results in Table~\ref{tab:tuning} indicate that SAEA performs best on the respective tuning instance of the objective function, followed by UEDA. The simple EA performs worst during tuning. In addition, the low standard deviation implies that SAEA is less sensitive to the tuned parameters, which may reasonably be attributed to the self-adaptive mechanism.

In terms of tuned parameters, the mutation rate for the EA is close to $1/n$, and the respective operator is bit-flip. This suggests that disturbing solutions as little as possible are preferable. The chosen recombination operator is a uniform crossover. This makes sense, as there is no obvious structure in the order of bits $z_i$ (i.e., the underlying data features, whose order is arbitrary) which would have to be preserved by using 1- or 2-point crossover.
The population size is the largest in the tested set of algorithms, which may be a reasonable measure to account for the relatively static behavior of the EA. 

The learning rate $\tau=0.048$ of the SAEA is set rather low, resulting in an average change of the mutation rate at only 3.8 \% per iteration. 
Considering that the initial mutation rate is at $1/n$, this means that it will probably remain fairly low throughout the complete run. 
Comparatively, the probability for changes of the operators is set rather high, leading to more frequent changes of the operator (which may either indicate that these operators have to change dynamically during the optimization run, or else, that the algorithm is not that sensitive to frequent changes of the operators).

The UEDA receives a rather high learning rate. This means, new probabilities $p^*_{z_i}$ estimated in an iteration contribute more to the respective probabilities in the next iteration than the probabilities employed in the current iteration of the UEDA. Seemingly, the UEDA requires changing the probabilities rather dynamically throughout each algorithm run.

Finally, the DCMA receives a population size that is slightly larger than the default at 13. The initial standard deviation is set rather low, and the margin parameter is set close to the suggested default $(\mu n)^{-1}$.

Only the more complex non-QC metaheuristics are tuned. The QAOA algorithm is not tuned before running it on the physical QC devices, since the accessibility of the system does not allow for the respective larger number of runs required for tuning.
Also, the metaheuristics, unlike QAOA, are not specifically developed for quadratic (or polynomial) objective functions and are hence, by default, less specialized to this problem class. 

\subsubsection{Validation experiment}
Validation is performed on the second of the two instances of the objective function.
The goal of this is an unbiased comparison, to avoid
a potential overfit of algorithm parameters to the tuning instance, although this is relatively unlikely given the rather rough tuning procedure.

For each algorithm, a maximum of 5000 objective function evaluations are allowed per independent run.
20 independent runs are performed for each metaheuristic algorithm (each run with different initial seeds).
  
Due to limitations in terms of computing access, QAOA does only 9 independent runs on the validation instance, with 58 to 79 iterations. Note, that each run takes three to four hours to complete (time-in-queue not included).
In comparison, a run of any of the metaheuristics takes 2 to 6 seconds, depending on hardware details (using no parallel computing resources).
CPLEX usually requires less than one second but makes use of multiple CPU cores, if available.

\subsubsection{Validation result}

\begin{figure}
	\centering
	\begin{subfigure}{.49\textwidth}
		\includegraphics[width=\textwidth]{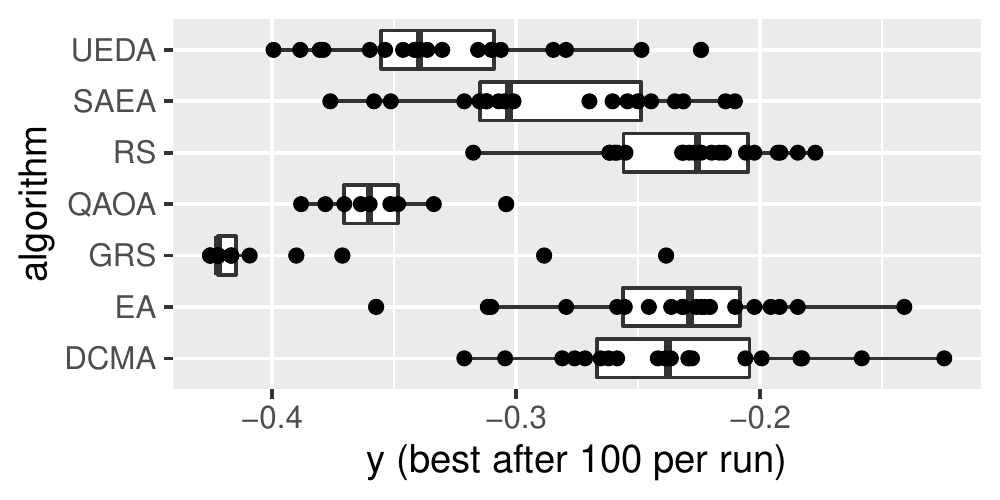}
	\end{subfigure}
	\begin{subfigure}{.49\textwidth}
		\includegraphics[width=\textwidth]{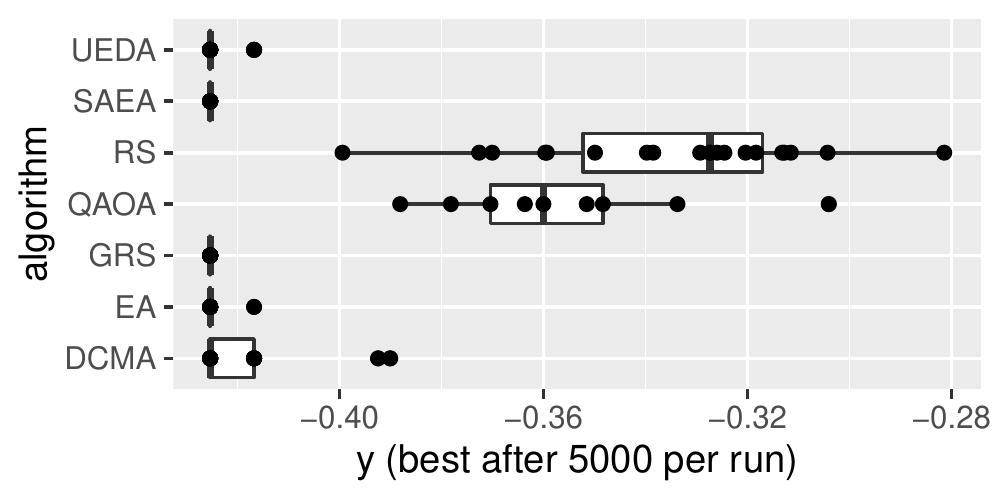}
	\end{subfigure}
	\begin{subfigure}{.49\textwidth}
		\includegraphics[width=\textwidth]{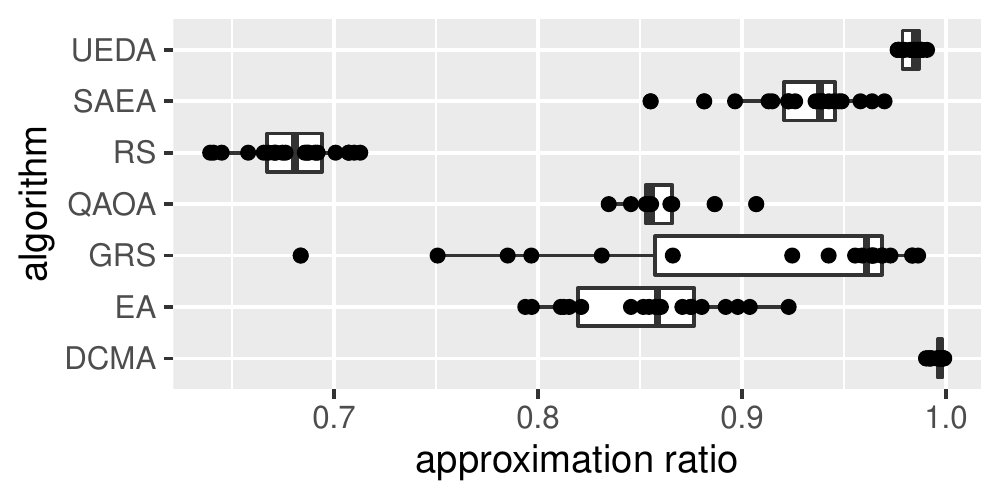}
	\end{subfigure}
	\begin{subfigure}{.49\textwidth}
		\includegraphics[width=\textwidth]{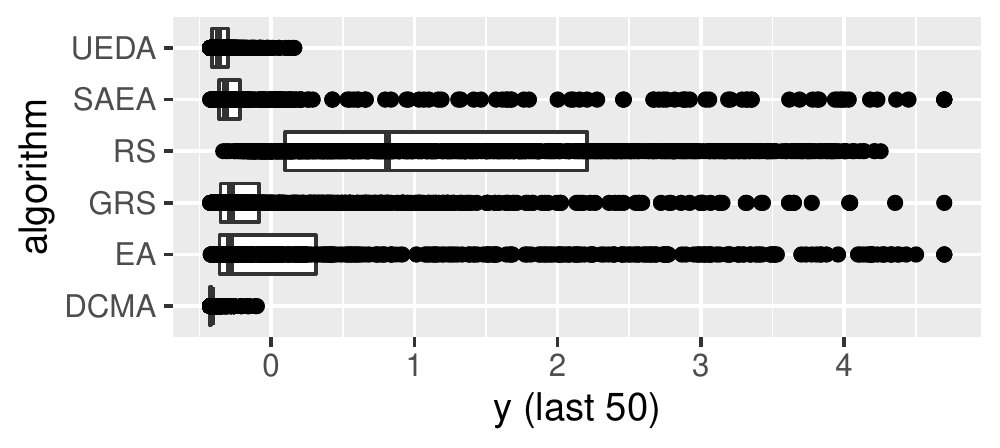}
	\end{subfigure}
	\caption{Top left: best found objective value after 100 objective function evaluations (OFE) per run. Top right: same after 5000 OFE.
    Note: QAOA is best overall, not counting OFE, hence the same in top left and top right.
    Bottom left: approximation ratio (of the last 50 of 5000 OFE in case of the non-quantum algorithms). Bottom right: raw objective function values of the last 50 OFE.
 }\label{fig:box}
\end{figure}

Figure~\ref{fig:box} shows some first results from the validation experiment. The upper two boxplots show the best solution founds (aggregated over the independent runs of each algorithm), early after just 100 classical objective function evaluations (OFE) and finally after 5000 OFE. Note that the numbers shown for QAOA are in each case recorded at the end of a terminated run.
Here, we see that while QAOA results are better than early results from each classical algorithm, the classical algorithms eventually find better results, given that enough OFE is invested.

\begin{figure}
    \centering
    \includegraphics[width=\textwidth]{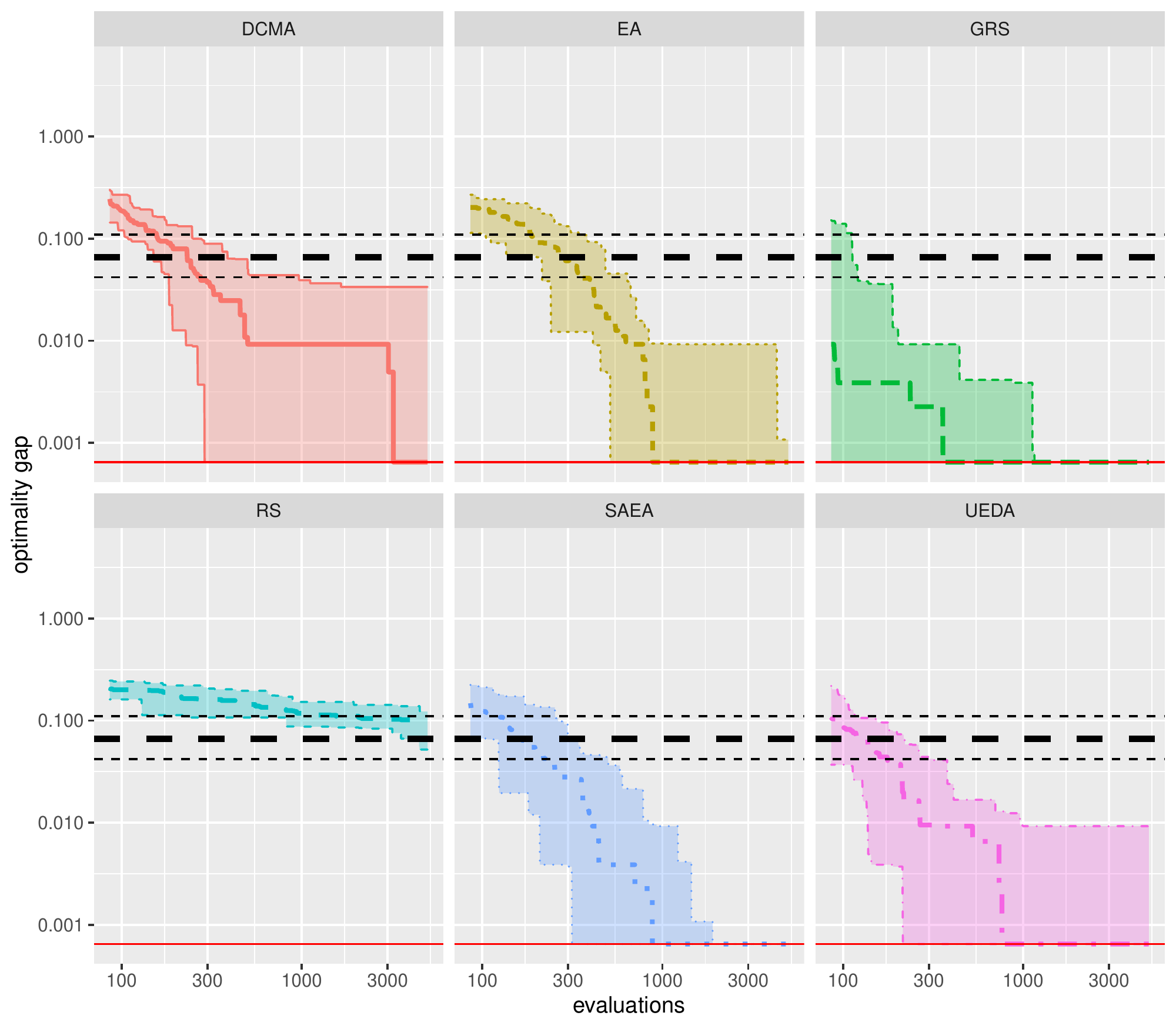}
    \caption{Optimality gap of the metaheuristics over the number of OFE (with log-scaled axes). Colored thick line is the median of each metaheuristic. The colored ribbon shows 5 and 95 percentiles. Black dashed lines show 5, 50, and 95 percentiles for QAOA. The thin red line at the bottom shows the global optimum value of the objective function (determined via CPLEX). The optimality gap is the difference of the best found objective function value to the optimum. The optimum is rounded down on the third digit to avoid infinite values. Hence, the global optimum is just below the 0.001 mark.}
    \label{fig:ybest}
\end{figure}

Between the classical algorithms, GRS performs best in terms of early performance. SAEA and GRS are the only algorithms to attain the global optimum in all runs within 5000 OFE, with GRS converging considerably faster.
This clear advantage of GRS is also visible in Fig.~\ref{fig:ybest}, and may indicate that the problem is `easy' to solve at least from the perspective of local search: only a few repeated local searches are sufficient to consistently determine the optimum.

The lower left of Fig.~\ref{fig:box} shows the approximation ratio (AE) after 5000 OFE. Note, that unlike the QAOA, AR values of the classical algorithms simply consider the last 50 evaluations of each run, weighting each observed candidate solution equally. The EA receives AR values comparable to QAOA, all other algorithms receive higher values. But note, that AR is not necessarily a comprehensive quality measure for the classical algorithms, since good performance (in terms of the best solution found) can be achieved by all four tested algorithms, despite significant differences in terms of AR. The reason here is that while final performances are equal, the variation of results differs a lot between the four algorithms, as the bottom right plot shows. Here, the EA and SAEA variants show much larger variations than the UEDA and DCMA. This is not (necessarily) a negative observation for EA and SAEA. In fact, larger variation within the sampled candidate solutions may imply a better ability to escape local optima.

\begin{figure}
    \centering
    \includegraphics[width=\textwidth]{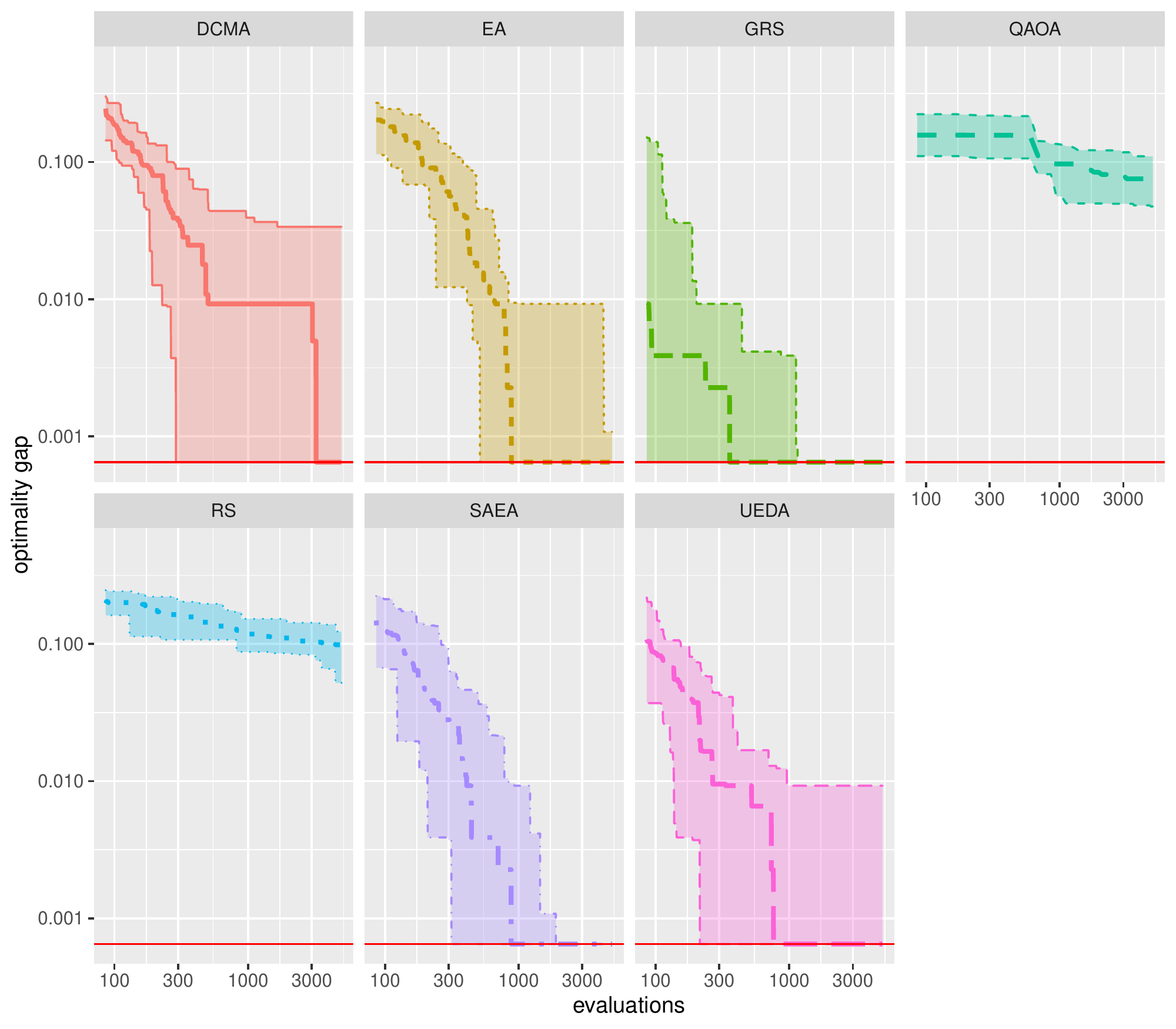}
    \caption{Optimality gap of the metaheuristics and QAOA over the number of OFE (log-scaled axes), similar to Figure~\ref{fig:ybest}, for the first 1000 OFE. Since QAOA does not consume OFE, it is instead reported against the number of QAOA iterations, where each QAOA iteration is (somewhat arbitrarily) valued at 86 OFE, to enable a comparison to the metaheuristics (i.e., scaled up to 5000 OFE). The optimality gap is the difference of the best found objective function value to the optimum. The optimum is rounded down on the third digit to avoid infinite values. Hence, the global optimum is just below the 0.001 mark.}
    \label{fig:ybestwq}
\end{figure}

As seen in Fig.~\ref{fig:ybest}, the metaheuristics require a few hundred of OFE to meet the performance of the final QAOA results. 
Since QAOA does not consume OFE, it is instead reported against the number of QAOA iterations in Fig.~\ref{fig:ybestwq}, where each QAOA iteration is (somewhat arbitrarily) valued at 86 OFE, to enable a visual comparison to the metaheuristic solvers.
The corresponding plot shows that QAOA itself starts with better objective function values than the competing algorithms, but after some initial gains, progresses only a little.
Only the simple RS is consistently outperformed by QAOA.

While QAOA (when run on actual QC hardware) achieves decent results fairly early, it is not yet able to compete with non-QC optimization algorithms. To determine whether this may change eventually will require a more in-depth investigation, that would have to make strong assumptions about, among others, how costs/effort of QC and non-QC algorithms are to be compared reasonably, and how they scale with aspects such as problem dimension or available parallel computing power.

\section{Discussion}

In this paper, the problem of selecting appropriate features for a supervised learning algorithm has been transformed to an optimization problem. This optimization problem has been solved with classical numerical methods as well as with quantum computing.

During the transformation of the feature selection problem, several ambiguities arise: One may choose between different dependency measures and choose whether the dependence between the features or the dependence between each feature and the target is more important. While the choice of the dependency measure depends on the data set considered, we found that in general a high value of $\phi$ is favored, which means, that the dependence between the target and the features is more important.

When comparing the performance of the optimization method (with a proper choice of the dependency measure and $\phi$), this method can compete with established methods like RFE or LASSO. However, all feature selection methods seem to be not able to find the global optimum, which could be obtained by brute force, for our small data sets. 

After confirming that the optimization method for feature selection works in principle, we explored different methods to solve the resulting quadratic binary optimization problem: A simple greedy search algorithm, a well-established commercial optimizer (IBM CPLEX), and a cutting-edge method of quantum computing.

For the quantum computing solution, we used the QAOA algorithm within the gate-based approach and used quantum simulators, as well as real quantum hardware available on the IBM quantum cloud. Whereas the solution of the small-size data sets merely served to demonstrate the feasibility, the solution of the 27-feature problem is undoubtedly more interesting. Although we were limited to $k<=3$ within QAOA, on real hardware we obtained approximation ratios above 0.80; well above the random sampling result. With the number of qubits available at the moment, the classical and meta-heuristical methods are not outperformed by quantum computing. An open question that will be addressed in future work is the scaling of both approaches, i.e. if and for which number of qubits there may be a quantum advantage.


\medskip 

{\bf Acknowledgements}
This work is funded by the Ministry of Economic Affairs, Labour and Tourism Baden-Württemberg in the frame of the Competence Center Quantum Computing Baden-Württemberg (project \lq QORA II\rq).

Useful discussions and the careful reading of the manuscript by PD Dr. Thomas Wellens, IAF, are greatly acknowledged.

{\bf Data availability}
The data that support the findings of this study are available upon reasonable request.
\medskip 

{\bf Declarations} The authors have no relevant financial or non-financial interests to disclose.


\newpage

\bibliography{Quellen}

\end{document}